\documentclass[prd,aps,twocolumn,a4paper,floatfix,nofootinbib,showpacs]{revtex4}

\usepackage{graphicx,psfrag}
\usepackage{mathrsfs}
\usepackage{amsmath,amsfonts,amssymb}
\usepackage{multirow}
\usepackage{comment}
\usepackage{placeins}
\usepackage[normalem]{ulem}

\newcommand{\be}{\begin{equation}}
\newcommand{\ee}{\end{equation}}
\newcommand{\beq}{\begin{equation}}
\newcommand{\eeq}{\end{equation}}
\newcommand{\bea}{\begin{eqnarray}}
\newcommand{\eea}{\end{eqnarray}}
\newcommand{\bel}{\begin{align}}
\newcommand{\eel}{\end{align}}

\def\GMc2{G M_{\odot} c^{-2}}

\usepackage{color}

\definecolor{cyan}{rgb}{0,0.9,0.9}
\definecolor{orange}{rgb}{0.9,0.5,0}
\definecolor{purple}{rgb}{0.8,0.4,0.8}
\definecolor{gray}{rgb}{0.8242,0.8242,0.8242}

\newcommand{\bamps}{\texttt{bamps} }
\newcommand{\e}[1]{$\cdot 10^{#1}$}

\begin{document}

\title{Solving 3D relativistic hydrodynamical problems\\ 
with WENO discontinuous Galerkin methods} 

\author{Marcus \surname{Bugner}$^1$}
\author{Tim \surname{Dietrich}$^1$}
\author{Sebastiano \surname{Bernuzzi}$^{2,3}$}
\author{Andreas \surname{Weyhausen}$^1$}
\author{Bernd \surname{Br\"ugmann}$^1$}

\affiliation{$^1$Theoretical Physics Institute, University of Jena, 07743
  Jena, Germany}
\affiliation{${}^2$Theoretical Astrophysics, California Institute of Technology, 
  1200 E California Blvd, Pasadena, California 91125, USA}
\affiliation{${}^3$DiFeST, University of Parma, I-43124 Parma, Italy}

\date{\today}

\begin{abstract} 
Discontinuous Galerkin (DG) methods coupled to WENO algorithms allow
high order convergence for smooth problems and for the simulation of
discontinuities and shocks. In this work, we investigate WENO-DG
algorithms in the context of numerical general relativity, in
particular for general relativistic hydrodynamics.
We implement the standard WENO method at different orders, 
a compact (simple) WENO scheme, as well as an alternative subcell
evolution algorithm.
To evaluate the performance of the different numerical schemes, we
study non-relativistic, special relativistic, and general relativistic
testbeds.
We present the first three-dimensional simulations of general
relativistic hydrodynamics, albeit for a fixed spacetime background,
within the framework of WENO-DG methods. 
The most important testbed is a single TOV-star in three dimensions,
showing that long term stable simulations of single isolated neutron
stars can be obtained with WENO-DG methods.

\end{abstract}


\maketitle

%%%%%%%%%%%%%%%%%%%%%%%%%%%%%%%%%%%%%%%%%%%%%%%%%%%%%%%%%%%%%%%%%%%%%%%%%%%%%%%%%%%%%%%%%%%%%%%%%%
\section{Introduction}
\label{sec:intro}
%%%%%%%%%%%%%%%%%%%%%%%%%%%%%%%%%%%%%%%%%%%%%%%%%%%%%%%%%%%%%%%%%%%%%%%%%%%%%%%%%%%%%%%%%%%%%%%%
Over the last decade simulations in numerical general relativity have seen a
tremendous improvement in accuracy and stability and have become an
important tool for the study of high energy and strong gravitational
field effects.  To date numerical simulations are the only possibility
to investigate complex astrophysical scenarios as e.g.~stellar
collapse~\cite{FryNew11} and coalescing binary neutron
stars~\cite{FabRas12}.  Although numerical simulations are in principle not
restricted by approximations beyond the numerical approximation, they are limited by the
finite accuracy of the particular discretization method.
Among the different methods to solve partial differential equations
like those of general relativity, the discontinuous
Galerkin (DG) method  has emerged in recent years as a
particularly successful general purpose paradigm
\cite{CanHusQua06,HesWar08,Kop09}.
It can be argued that the DG method, more explicitly the DG finite
element method or DG spectral element method, subsumes and combines
several of the key advantages of traditional finite element and
finite volume methods, e.g.\ \cite{HesWar08}.
In particular, the {\em discontinuous} Galerkin method works with
element-local stencils, which is a great advantage for parallelization
and the construction of complicated grids. Furthermore,
DG methods offer easy access to
$hp$-adaptivity \cite{KarShe05}, where both the size of the computational
elements (or cells) and the order of the polynomial approximation
within each element can be adapted to the problem.  For smooth
solutions, DG methods approach the optimal order of exponential
convergence of pseudospectral methods on multiple patches. In fact,
certain DG methods are equivalent to pseudospectral methods with a
specific penalty method for the patch boundaries \cite{HesGotGot07}.
For non-smooth solutions, low order elements have been combined with
various HRSC schemes, for example in
the form of WENO DG methods \cite{QiuShu05}.

In this work we consider the application of DG methods to simulations
in numerical general relativity coupled to general relativistic
hydrodynamics (GRHD). Concretely, the goal is to compute the numerical
evolution of spacetimes containing neutron stars. The governing
differential equations are the time-dependent, non-linear Einstein
field equations for the spacetime geometry coupled to a relativistic
fluid model. 
Most relativistic hydrodynamics simulations are based on the ``Valencia
formulation'', in which the matter field evolution is given in a
conservative form~\cite{Fon07}. 

Among the numerous numerical studies carried out in the field, most
have been performed using finite difference (FD) and finite volume (FV) methods,
with significant success. 
For the geometry (including black holes), high-order finite
differencing is the rule, often 4th to 8th order finite differences in
space for structured adaptive mesh refinement (AMR),
e.g.\ \cite{BruGonHan06,SchDieDor06}. The matter part allows the
formation of strong relativistic shocks, and a variety of finite
volume (or finite difference) HRSC schemes have been developed
\cite{Fon07,RezZan13}.  For smooth solutions, pseudospectral methods
have been very successful \cite{SPEC,HilWeyBru15,Tic09,GraNov09}.
Recently, a convergence order of $\sim$ 3 was observed for high order
matter formulations in~\cite{RadRezGal13,RadRezGal14,RadRezGal15}. 

DG methods for numerical relativity offer the usual list of attractive
features. In particular, one goal would be to combine high-order,
smooth regions with lower-order regions containing shocks. Compared
to AMR with large, overlapping finite difference stencils, the DG spectral
element method is more easily and more efficiently parallelizable, while 
still allowing high-order approximations.
However, there remain several open issues with regard to DG methods in
numerical relativity. Some issues are known but unresolved, some have
simply not been investigated yet.

The evolution equations of numerical relativity are a coupled system
for the geometry (the metric variables) and the matter variables. 
While the matter equations are naturally given in a flux
form \cite{Fon07}, this is not the case for the geometry. 
Since a typical DG method starts with a flux-balance law, it is in
principle straightforward to design a method for the matter part. 
On the other hand, for the geometric part one should either recast the
equations in a hyperbolic flux form, or suggest less standard methods.

There have been essentially only three major efforts to employ DG
methods for general relativity and/or GRHD. In \cite{Zum09}, Zumbusch
gives the first and so far only example for a complete DG method for
the ($3+1$)-dimensional (short 3D) Einstein equations in
vacuum. Discussed is a space-time DG scheme mostly in the context of
linearized equations and in a specific gauge, but the scheme also
handles non-linearities. So far there has not been an astrophysics
application, say involving black holes or neutron stars.
In \cite{BroDieFie12}, Brown et al.\ discuss a DG method
for the so-called BSSN formulation of the vacuum equations,
mostly with 1D examples. The BSSN equations are not in flux-form, but
the various non-linearities and second derivatives are successfully
dealt with on a case by case basis.
And in \cite{RadRez11}, Radice and Rezzolla give a general
discussion of the DG method for the matter equations in the standard
flux form, without including the equation for the geometry. They
present a working 1D implementation for general relativistic matter in
spherical symmetry.
In addition, there has been work on special relativistic hydrodynamics
(SRHD).  Zhao and Tang \cite{ZhaTan13} were the first to apply the
WENO-DG method of \cite{QiuShu05} to a variety of 1D and 2D test cases
in SRHD, and the method turned out to be robust and reliable in
capturing shocks.

The concrete target of the present work is to model a single
stationary neutron star (a TOV star \cite{Tol39,OppVol39a}) in 3D,
although we perform a variety of tests in 1D and 2D as well.
The TOV star is computed in the Cowling approximation, which
simplifies the problem by assuming that the geometry may be curved but
does not depend on time, which in turn is compatible with the
stationarity of the TOV star. The numerical evolution of the matter
variables for fixed metric is a standard approach that still allows to
test key features of the hydrodynamics, including the treatment of the
non-differentiable density at the surface of the star.
We leave the coupling to a dynamic geometry to future work. 

In preparation for the simulations in full, 3D GRHD, we test the
Runge-Kutta DG (RKDG) method coupled to a variety of WENO
reconstructions for the equations of general relativistic
hydrodynamics~\cite{Fon07}. We reproduce the non-relativistic standard
results~\cite{QiuShu05,ZhoShu13} as well as some of the special
relativistic test cases of~\cite{ZhaTan13} for a third and a fifth
order method, WENO3 and WENO5. We extend \cite{ZhaTan13} by also
considering WENO-Z \cite{BorCarCos08} and the simple WENO limiters of
\cite{ZhoShu13}. Finally, we present the first application of RKDG
WENO methods to a 3D TOV star in the Cowling approximation.
The numerical experiments are implemented in the new \bamps
code~\cite{HilWeyBru15} for spectral element methods. We import some methods 
from an existing full-featured finite difference AMR code for 3D GRHD, BAM
\cite{ThiBerBru11,DieBerUje15,BerNagThi12,DieMolJoh15,BruGonHan06}.

When researching the available HRSC methods for DG, there is one issue
related to shock resolution and efficiency that is well-known but that
does not always appear to receive the attention it deserves.
For FD or FV methods with HRSC, shocks are resolved within a few cell
widths, which means within a few grid points. For DG methods, the
standard approach is to employ WENO reconstruction based on cell
averages \cite{QiuShu05}.  In such WENO-DG methods, shocks are again
resolved within a few cell widths, but each cell now contains $p$
points (for polynomials of order $p-1$). The WENO3 stencil involves 3
cells and $3p$ points, and the WENO5 stencil involves 5 cells and $5p$
points.
Effectively, the high resolution within each cell (the ``subcell resolution'') 
is lost if only the cell averages are used.
For practical implementations a rough estimate is therefore that such
WENO-DG methods could require about $27$ or $125$ times more resources for shock resolution in
3D than comparable FD or FV methods (these factors vary with the
actual implementation).

For the evaluation of DG methods for GRHD it matters whether such methods
are competitive to existing FD/FV methods in terms of
efficiency. Hence we consider the following measures aimed at handling
the comparatively low efficiency of cell-averaged WENO-DG methods. A
common strategy is to limit the application of the WENO scheme to
only those cells that need it, and there has been quite some work on
so-called ``troubled cell indicators'' \cite{QiuShu05a}. 

A recent development are the so-called ``simple'' WENO methods of
\cite{ZhoShu13}, which effectively construct a compact stencil for
high-order WENO methods. For example, the fifth-order WENO method is
constructed from only 3 instead of 5 cells, using the high-order
information from the nearest neighbor cells to obtain fifth order.
This leads to significant savings, but the method has not been widely
tested yet. We include the compact/simple WENO method in our tests
and report on some differences to the standard WENO method, in
particular in 3D.

Another important development is a hybrid approach
\cite{DumZanLou14,ZanFamDum15,SonMun14,HueCasPer12}, 
which replaces troubled cells by an equidistant
subgrid and applies FV shock capturing on these grids. This approach maintains
the subcell resolution of FV methods, but increases the complexity of
the implementation since two types of grids and special grid transfer
operators are required. In our case the method is appealing because a
full-featured FD implementation is already
available~\cite{ThiBerBru11,DieBerUje15}. If successful, the strategy would be to 
construct a high-order DG method for regions where the solution is smooth, but
to rely on established FD methods near shocks. 

HRSC for DG comes at a cost since WENO-DG as well as the hybrid FD-DG
method break the cell-locality of the basic DG method.  We consider
both methods here to gain some insight into their relative merit.

In Sec.~\ref{sec:NumMeth}, we introduce the DG method for 3D
flux-balance laws, specify the equations of relativistic
hydrodynamics, and discuss the WENO-DG and FD-DG methods.
In Sec.~\ref{sec:numerics}, we summarize the numerical implementation.
As basic tests we consider the advection equation and the Burgers equation
in 1D in Sec.~\ref{sec:simple_tests}, while 1D and 2D tests for SRHD are presented in
Sec.~\ref{sec:SRHD}.
The main results concern the evolution of a TOV star in Sec.~\ref{sec:GRHDC}.
We conclude in Sec.~\ref{sec:Conclusion}.
For completeness, we collect some relevant details of the basic 1D DG method in
App.~\ref{sec:appendixDG}.

Throughout the article dimensionless units are used, i.e.~we set 
$c=G=M_\odot=1$.  
We denote spacetime indices
by~$a, b, \ldots$ and indices over space dimensions by~$i,j,\ldots$.

%%%%%%%%%%%%%%%%%%%%%%%%%%%%%%%%%%%%%%%%%%%%%%%%%%%%%%%%%%%%%%%%%%%%%%%%%%%%%%%%%%%%%%%%%%%%%%%%%%
%%%%%%%%%%%%%%%%%%%%%%%%%%%%%%%%%%%%%%%%%%%%%%%%%%%%%%%%%%%%%%%%%%%%%%%%%%%%%%%%%%%%%%%%%%%%%%%%%%
\section{Methods}
\label{sec:NumMeth}
%%%%%%%%%%%%%%%%%%%%%%%%%%%%%%%%%%%%%%%%%%%%%%%%%%%%%%%%%%%%%%%%%%%%%%%%%%%%%%%%%%%%%%%%%%%%%%%%%%
%%%%%%%%%%%%%%%%%%%%%%%%%%%%%%%%%%%%%%%%%%%%%%%%%%%%%%%%%%%%%%%%%%%%%%%%%%%%%%%%%%%%%%%%%%%%%%%%%%

%%%%%%%%%%%%%%%%%%%%%%%%%%%%%%%%%%%%%%%%%%%%%%%%%%%%%%%%%%%%%%%%%%%%%%%%%%%%%%%%%%%%%%%%%%%%%%%%%%
\subsection{Discontinuous Galerkin method}
\label{sec:DG}
%%%%%%%%%%%%%%%%%%%%%%%%%%%%%%%%%%%%%%%%%%%%%%%%%%%%%%%%%%%%%%%%%%%%%%%%%%%%%%%%%%%%%%%%%%%%%%%%%%

The hydrodynamical equations governing the time evolution of the matter fields  
can be cast as a non-linear conservation law for a vector of variables,
$\mathbf{u}(x,t)$, depending on time $t$ and position $x\in\mathbb{R}^3$. 
The conservation law is given by
\begin{equation}
  \partial_t \mathbf{u} + \partial_i \mathbf{f}^i(\mathbf{u}) = \mathbf{S} \quad ,
  \label{eq:con_law}
\end{equation}
with the sources $\mathbf{S}$ and the fluxes $\mathbf{f}^i$.
We summarize some of the relevant aspects of the DG method for
conservation laws of scalar function on $\mathbb{R}$ in Appendix
\ref{sec:appendixDG}, while simply stating the key equations for
vector-valued functions on $\mathbb{R}^n$ here,
cmp.\ \cite{HesWar08,ZhaTan13}.
  
We consider a partition of $\mathbb{R}^n$ into cells $I_j$, $x \in I_j$,  
and define the finite dimensional approximation space
\begin{equation}
  V^N := \left\{ v: v(x)|_{I_j} \in \mathbb{P}^N(I_j) \right\}
  \label{eq:sol_space}
\end{equation}
with $\mathbb{P}^N(I_j)$ denoting the finite dimensional space of polynomials on $I_j$ of degree 
at most $N$. As in most standard applications, we set the polynomial order $N$ as a constant over
the whole partition. To deduce a DG scheme from Eq.~\eqref{eq:con_law}, we want to find a 
function $\mathbf{u_n}(x)$ for which the weak form
\begin{align}
   & \int_{I_j} v \partial_t \mathbf{u_n}\ \text{d}V + 
     \int_{\partial I_j} \mathbf{f}^{i}(\mathbf{u_n}) v n_i \ \text{d}S \notag\\
   &-\int_{I_j} \mathbf{f}^i(\mathbf{u_n}) \partial_i v \ \text{d}V = 
     \int_{I_j} \mathbf{S} v \ \text{d}V
 \label{eq:weak_form}
\end{align}
holds for all $v \in V^N$. For simplicity, we denote the approximate/numerical solution $\mathbf{u_n}(x)$
as $\mathbf{u}(x)$ in the following. An important advantage of the DG-scheme is that $v$ does not 
need to be continuous at the cell boundaries. Therefore, no unambiguous definition of the fluxes 
at cell boundaries entering Eq.~\eqref{eq:weak_form} exists. To overcome this issue, we 
introduce the numerical fluxes $\mathbf{f^{*}}^i(\mathbf{u_-},\mathbf{u_+})$,
which depend on the inside/outside cell limited value of $\mathbf{u}$ at the boundary,
$\mathbf{u_-}$ and $\mathbf{u_+}$, and reproduce the original flux if $\mathbf{u}$ 
is continuous. A simple example of a numerical flux with this property is 
the local \textsc{Lax}-\textsc{Friedrich} (LLF) flux 
\begin{align}
  \mathbf{f^*}^i(\mathbf{u_-},\mathbf{u_+})n_i = \frac{1}{2} \left[
  \mathbf{f}^i(\mathbf{u_-})n_i + \mathbf{f}^i(\mathbf{u_+})n_i -\lambda \left(
  \mathbf{u_+} - \mathbf{u_-}
  \right)\right] \quad ,
\label{eq:LFF}
\end{align}
where $\lambda$ denotes the maximum absolute eigenvalue of the Jacobian 
$\partial (\mathbf{f}^in_i)/\partial \mathbf{u}$. We use an LLF algorithm throughout 
this article. 
Writing out the numerical solution $\mathbf{u}(x,t)$ as an element of $V^N$
explicitly,
\begin{align}
  \mathbf{u}|_{I_j}(x,t) &= \sum_{k=0}^{N}
  \mathbf{\hat{u}}_k(t) v^{k}(x)
\label{eq:Ansatz_u}
\end{align}
and $v^{k}$ a basis of $\mathbb{P}^N(I_j)$, allows to recast~\eqref{eq:weak_form} as
an algebraic equation for the unknown time derivatives $\partial_t \mathbf{\hat{u}}_{k}$.
To evolve these coefficients in time, we use an explicit fourth order Runge-Kutta method.
More details on the actual implementation are given in Sec.~\ref{sec:numerics}.  
        
%%%%%%%%%%%%%%%%%%%%%%%%%%%%%%%%%%%%%%%%%%%%%%%%%%%%%%%%%%%%%%%%%%%%%%%%%%%%%%%%%%%%%%%%%%%%%%%%%%
\subsection{Relativistic hydrodynamics}
%%%%%%%%%%%%%%%%%%%%%%%%%%%%%%%%%%%%%%%%%%%%%%%%%%%%%%%%%%%%%%%%%%%%%%%%%%%%%%%%%%%%%%%%%%%%%%%%%%

Although we are working in cowling approximation, i.e.~keeping the metric fixed,
the matter fields are evolved dynamically on a curved spacetime background.
We want to recast briefly the important equations and methods necessary to 
solve the general relativistic hydrodynamical equations; 
special relativity can be easily obtained by choosing flat spacetime.
  
%%%%%%%%%%%%%%%%%%%%%%%%%%%%%%%%%%%%%%%%%%%%%%%%%%%%%%%%%%%%%%%%%%%%%%%%%%%%%%%%%%%%%%%%%%%%%%%%%%
\subsubsection{3+1-decomposition}

Although we assume the spacetime to be fixed, we have to recast it in a suitable 
form for dynamical evolutions. This can be done with the help of a~$3+1$ decomposition~\cite{ArnDesMis04,Yor79}
(see~\cite{Alc08,BauSha10,Gou12} for textbook introductions) in which
the four-dimensional spacetime metric is rewritten as
\begin{equation}
  {\rm d}s^2  =  -\alpha^2 {\rm d}t^2 + \gamma_{ij} ({\rm d}x^i + \beta^i{\rm d}t)({\rm d}x^j + \beta^j {\rm d}t), \label{eq:4metric}
\end{equation}
where~$\alpha$ is the lapse function,~$\beta^i$ the shift vector, and~$\gamma_{ij}$ the spatial metric. 
In case of a flat spacetime $\alpha=1, \beta^i=0, \gamma_{ij}= \delta_{ij}$ employing Cartesian coordinates. 
Einstein's field equations split into two sets, the constraint equations and the evolution equations.
For our single neutron stars tests, we recast the Tolman-Oppenheimer-Volkoff (TOV)-equation~\cite{Tol39,OppVol39a} in 3+1-form and solve it to obtain an
ordinary differential equations. 
In addition to the 3+1-split we perform a conformal transformation of the spatial metric,
\begin{align}
  \gamma_{ij} = \psi^4 \bar \gamma_{ij},
\end{align}
where~$\psi$ is the conformal factor and $\bar \gamma_{ij}$ the conformally related metric. 

  %%%%%%%%%%%%%%%%%%%%%%%%%%%%%%%%%%%%%%%%%%%%%%%%%%%%%%%%%%%%%%%%%%%%%%%%%%%%%%%%%%%%%%%%%%%%%%%%%%
\subsubsection{Hydrodynamic equations}
\label{sec:hydro}

According to Eq.~\eqref{eq:con_law} we denote the state vector collecting the
conserved variables as $\mathbf{u}$, while $\mathbf{f}^{i}(\mathbf{u})$ are hydrodynamical fluxes, 
and $\mathbf{S}$ the source terms. The fluxes and the sources depend in general on the metric and matter fields.
The conserved variables are $\mathbf{u} = \sqrt{\gamma}(D,\,S_k,\,\tau)$, 
and denote respectively the rest-mass density ($D$), the momentum
density ($S_k$), and an internal energy ($\tau$) measured by the Eulerian
observer given by the particular spacetime foliation. 
$\gamma=\det\gamma_{ij}$ is the determinant of the spatial three-metric. 
The conserved variables $\mathbf{u}$ can be reconstructed from the primitive variables 
$\mathbf{w}=(\rho,v^i,\epsilon, p)$, i.e.~rest-mass density,
3-velocity measured by the Eulerian observer, internal energy and pressure of the fluid 
by the following equations: 
\begin{subequations}
\begin{eqnarray}
  D    &  = & W \rho, \label{eq:cons1} \\
  S_k  &  = & W^2 \rho h v_k, \label{eq:cons2} \\
  \tau &  = & (W^2 \rho h - p ) - D, \label{eq:cons3}
\end{eqnarray}
\end{subequations}
where $W$ is the Lorentz factor, $W=1/\sqrt{1-v_i v^i}$ and $h$ is the 
specific enthalpy $h=1+\epsilon + p/\rho$.

To close the system an equation of state (EOS) $p=P(\rho,\epsilon)$
is needed. In this work, we use a simple polytropic 
\begin{equation}
  P(\rho) = K \rho^\Gamma \label{eq:EOS_poly}
\end{equation}
or an ideal gas EOS of the form
\begin{equation}
  P(\rho,\epsilon) = (\Gamma-1) \rho \epsilon \label{eq:EOS_ideal}.
\end{equation}
The particular implementation of the hydrodynamical equations follows~\cite{Fon07,ThiBerBru11}.
  
However, due to the special choice of the background metric, 
the flux and source terms simplify dramatically by setting 
$\bar{\gamma}_{ij}=\delta_{ij}$ and $\beta^i=0$ in all our examples. 

%%%%%%%%%%%%%%%%%%%%%%%%%%%%%%%%%%%%%%%%%%%%%%%%%%%%%%%%%%%%%%%%%%%%%%%%%%%%%%%%%%%%%%%%%%%%%%%%%%
\subsubsection{Primitive recovery and atmosphere treatment}

We evolve the conservative variables $\mathbf{u}$ by constructing the fluxes and source terms for  
every time slice. While $\mathbf{f}^i$ and $\mathbf{S}$ both contain the primitive variables $\mathbf{w}$
we have to recover those from the conservatives. 
The inverse relations of \eqref{eq:cons1}-\eqref{eq:cons3} are given by:
\begin{eqnarray}
  \rho     & = & \frac{D}{W}, \label{eq:prim1} \\
  v^i      & = & \frac{S^i}{\tau+D+p}, \label{eq:prim2}\\
  \epsilon & = & \frac{\sqrt{(\tau+p+D)^2-S^2}-W p -D}{D},\label{eq:prim3}
\end{eqnarray}
with $W=(\tau+p+D)/\sqrt{(\tau+p+D)^2-S^2}$ and $S^2=S_i S^i$. 
To make use of \eqref{eq:prim1}-\eqref{eq:prim3}, we have to determine the pressure $p$.

The explicit primitive reconstruction goes as follows. 
First, we try to recover the primitive variables 
for the full equation of state including thermal components $p=P(\rho,\epsilon)$. 
For this reason a Newton-Raphson method is employed to compute the pressure $p$. If the method does 
not converge to the desired accuracy a cold equation of state $p=p(\rho)$ is used and we try to find with a 
Newton-Raphson method the density $\rho$.

As in most general relativistic hydrodynamic codes, we have to 
include an artificial atmosphere to solve the problem of fluid-vacuum interfaces. 
This allows long term stable and robust numerical simulations~\cite{FonMilSue00,BaiHawMon04,Dimmelmeier:2002bk}. 
The atmosphere $\rho_{\rm atm}$ is computed according to 
\begin{equation}
  \rho_{\rm atm} = f_{\rm atm} \cdot \text{max}[\rho(t=0)].
\end{equation}
Whenever a point falls below the atmosphere threshold $\rho_{\rm thr} = f_{\rm thr}\cdot \rho_{\rm atm}$  
during the evolution or the primitive reconstruction, it is set to the atmosphere value. 
  
%%%%%%%%%%%%%%%%%%%%%%%%%%%%%%%%%%%%%%%%%%%%%%%%%%%%%%%%%%%%%%%%%%%%%%%%%%%%%%%%%%%%%%%%%%%%%%%%
\subsection{WENO reconstruction methods}
\label{sec:WENO_method}
%%%%%%%%%%%%%%%%%%%%%%%%%%%%%%%%%%%%%%%%%%%%%%%%%%%%%%%%%%%%%%%%%%%%%%%%%%%%%%%%%%%%%%%%%%%%%%%%%%

\begin{figure*}[t]
  \centering
  \includegraphics[width=\textwidth]{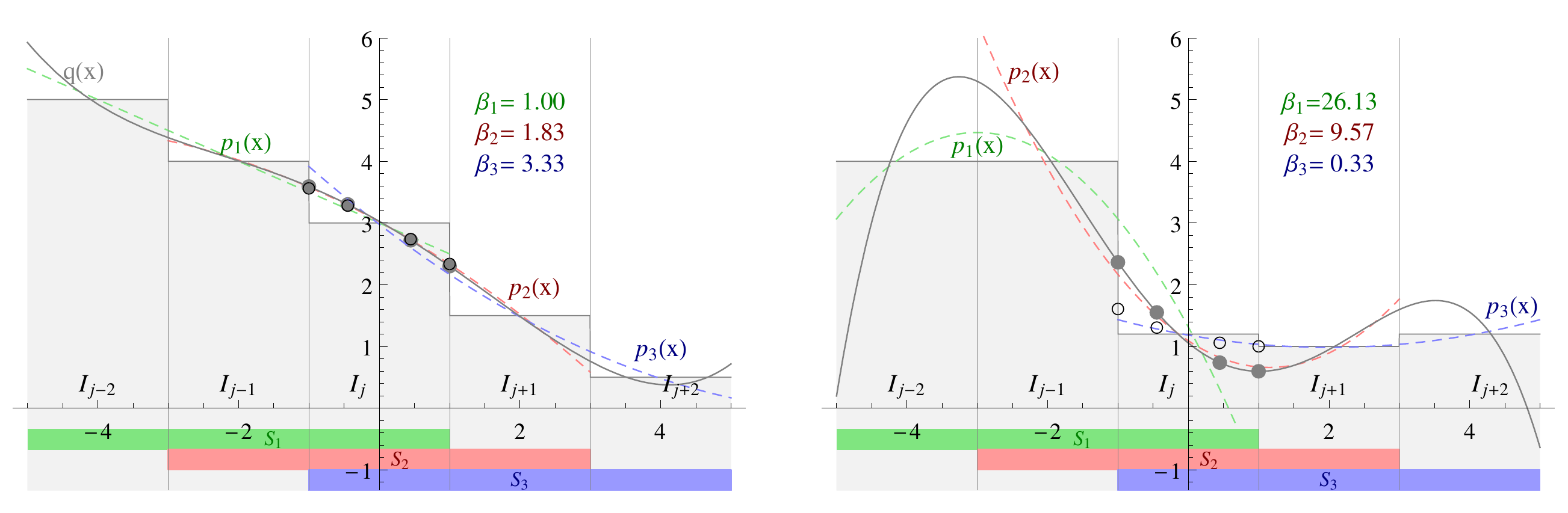}
  \caption{The WENO-5 ($w=2$) methodology applied in a smooth case (left figure) and a shock case (right figure). The values in the interval
  $x \in I_j = [-1,1]$ are to be reconstructed from the five grid patch averages $u_{j-2},u_{j-1},u_{j},u_{j+1},u_{j+2}$. The three stencils 
  $S_1, S_2, S_3$ are created as a clustering of three grid patches each with the corresponding approximating polynomial
  $p_1(x),p_2(x),p_3(x)$. Another higher order polynomial $q(x)$ can be found from employing all five averages. Following the strategy as described
  in~\ref{sec:WENO_method}, the smoothness indicators $\beta_i$ are calculated for each stencil. A large $\beta_i$ indicates non-smoothness
  of the corresponding polynomial $p_i$, which leads to a minor contribution of the stencil $S_i$ for the reconstruction. In the shock 
  case, the reconstructed point values (empty black circles) lie very close to the smoothest polynomial $p_3$, whereas in the smooth case
  all three approximating polynomials are taken into account almost equally, so that the reconstruction is very close to the 5th order
  polynomial $q$ (filled gray circles).}
  \label{fig:1DWENO}
\end{figure*}

As a next step, we explain how to avoid oscillations and unphysical behavior caused by the Gibbs phenomenon.
For this purpose we locate discontinuities and oscillations with the troubled cell indicator described in Sec.~\ref{sec:troubled_cell_indicator}
and apply a WENO limiter reconstruction~\cite{QiuShu05,ZhaTan13}.
We introduce three different WENO reconstruction methods, the standard WENO approach (Sec.~\ref{sec:standard_WENO}), 
the simple WENO algorithm~\cite{ZhoShu13} based on compact stencils (Sec.~\ref{sec:simple_WENO}), 
and a WENO algorithm based on a subcell evolution (Sec.~\ref{sec:subcell_method}). 

%%%%%%%%%%%%%%%%%%%%%%%%%%%%%%%%%%%%%%%%%%%%%%%%%%%%%%%%%%%%%%%%%%%%%%%%%%%%%%%%%%%%%%%%%%%%%%%%
\subsubsection{Troubled cell indication}
\label{sec:troubled_cell_indicator}

Given the coefficients of the numerical solution $\mathbf{\hat{u}}_{p}(t)$ at time $t$, 
we can calculate the average of the polynomial
$\mathbf{u}(x,t)$ over the grid patches $I_j = [a_j,b_j]$:
\begin{align}
  \mathbf{u_j} := \frac{1}{\Delta x} \int_{a_j}^{b_j} \mathbf{\hat{u}}_{p} \phi^{p}(x) \ dx = \frac{1}{2} \int_{-1}^{1} \mathbf{\hat{u}}_{p} \ell^{p}(\xi) \ d\xi \ .
\label{eq:WENOaverages}
\end{align}
We further denote the boundary values of $\mathbf{u}$ as
\begin{align}
  \mathbf{u}_j^{-} := \mathbf{u}(a_j), \qquad \mathbf{u}_j^{+} := \mathbf{u}(b_j)
\label{eq:WENOboundaries}
\end{align}
and define the four differences:
\begin{eqnarray}
  \mathbf{\tilde{u}_j^{-}} :=& \mathbf{u_j} - \mathbf{u}_j^{-}, \qquad \mathbf{\tilde{u}_j^{+}} &:= \mathbf{u}_j^{+} - \mathbf{u_j}\\
  \Delta_{-}\mathbf{u}     :=& \mathbf{u_j} - \mathbf{u_{j-1}}, \qquad \Delta_{+}\mathbf{u}     &:= \mathbf{u_{j+1}} - \mathbf{u_j}
\label{eq:TVBdifferences}
\end{eqnarray}
We also introduce the minmod function
\begin{align}
  &\text{minmod}(x_1,x_2,...,x_n) = \notag\\
  &\begin{cases}
   s \cdot \min_{1 \leq j \leq n} |x_j| & \text{if } \text{sign}(x_1) =  ... = \text{sign}(x_n) =: s \\
   0 & \text{otherwise}
  \end{cases}
\label{eq:minmod}
\end{align}
and the modified TVB minmod function
\begin{align}
  &\text{minmod}_\text{TVB}(x_1,x_2,...,x_n) = \notag\\
  &\begin{cases}
  a_1 & \text{if } |a_1| \leq M \left(\max_j \Delta x_j\right)^2 \\
  \text{minmod}(x_1,x_2,...,x_n) & \text{otherwise}
  \end{cases}
\label{eq:TVBminmod}
\end{align}
Our troubled cell indicator marks a grid patch as troubled, if
\begin{align}
  &\text{minmod}_\text{TVB}\left(\left(\tilde{u}_j^{-}\right)^k , \Delta_{-}u^k, \Delta_{+}u^k \right) \neq \left(\tilde{u}_j^{-}\right)^k \quad \text{ or } \notag\\ 
  &\text{minmod}_\text{TVB}\left(\left(\tilde{u}_j^{+}\right)^k , \Delta_{-}u^k, \Delta_{+}u^k \right) \neq \left(\tilde{u}_j^{+}\right)^k
\label{eq:tc_ind}
\end{align}
for at least one component $k$. This is exemplary for a situation, in which a component of $\mathbf{u}$ is not monotonous (because the arguments 
of minmod differ in sign) or its gradient inside a patch is larger, than that of the neighboring patches (shock inside the cell).

In the case of multiple dimensions, we perform the 1D troubled cell indication in every direction. A cell is marked as
troubled, if at least one of these indications results in a troubled state. To apply the 1D algorithm, the boundary values used in \eqref{eq:WENOboundaries}
have to be modified, since the cell boundaries are not longer single points, but lines or surfaces. Therefore, we redefine $\mathbf{u_j^{\pm}}$ by the 
boundary averages, i.e. for a 3D cell $I = [a_j, b_j]\times[a_k, b_k]\times[a_l, b_l]$ in $x$-direction
\begin{align}
  \mathbf{u_{jkl}^{-}} &:= \frac{1}{\Delta y \Delta z} \int_{a_k}^{b_k}\int_{a_l}^{b_l}\mathbf{\hat{u}}_{pqr} \phi^{p}(a_j) \phi^{q}(y) \phi^{r}(z) \ \text{d}y \text{d}z , 
  \notag\\
  \mathbf{u_{jkl}^{+}} &:= \frac{1}{\Delta y \Delta z} \int_{a_k}^{b_k}\int_{a_l}^{b_l}\mathbf{\hat{u}}_{pqr} \phi^{p}(b_j) \phi^{q}(y) \phi^{r}(z) \ \text{d}y \text{d}z .
  \label{eq:2DWENOaverages}
\end{align}

%%%%%%%%%%%%%%%%%%%%%%%%%%%%%%%%%%%%%%%%%%%%%%%%%%%%%%%%%%%%%%%%%%%%%%%%%%%%%%%%%%%%%%%%%%%%%%%%
\subsubsection{Standard WENO reconstruction}
\label{sec:standard_WENO}

In a standard WENO method of order $2w+1$, we construct $w+1$ stencils $S_i$ around $I_j$, each as an aggregation of $w+1$ grid patches: 
$S_i = (I_{j-w+i},I_{j-w+i+1},...,I_{j+i}), 0 \leq i \leq w$. In Fig.~\ref{fig:1DWENO} this partitioning is shown for $w=2$. For each stencil,
we construct a $w$-th order polynomial $\mathbf{p}_i$, which has the same average as the numerical solution $\mathbf{u}$ over each 
grid patch in the stencil. That means solving the system
\begin{align}
  \mathbf{u_k} = \frac{1}{\Delta x}\int_{I_k} \mathbf{p}_i(x) \ \text{d}x, \qquad \text{for all } I_k \in S_i
\label{eq:WENOpcondition}
\end{align}
for the $w+1$ coefficients of each component of $\mathbf{p}_i$. 
Similarly, we construct a $2w$-th order polynomial $\mathbf{q}$ fulfilling
\begin{align}
  \mathbf{u_k} = \frac{1}{\Delta x} \int_{I_k} \mathbf{q}(x) \ \text{d}x, \qquad \text{for all } I_k \in S,
\label{eq:WENOqcondition}
\end{align}
with $S := \cup_{i} S_i$ being the large stencil over all $2w+1$ grid patches.
The fundamental concept is to approximate the solution in $[-1,1]$ as a linear combination of the $\mathbf{p}_i$, which should give the same result as the
higher order approximation $\mathbf{q}$ in smooth regions.
This condition defines the linear (or ideal) weights $\gamma_i$ satisfying
\begin{align}
  \mathbf{q}(x) = \sum_{i=1}^{w+1} \gamma_i(x) \mathbf{p}_i(x).
\label{eq:WENOgcondition}
\end{align}
We emphasize that the $\gamma_i$ depend on the point $x$ where the approximation should hold. 
It is remarkable that although both sides of Eq.~\eqref{eq:WENOgcondition} depend intrinsically on the $2w+1$ averages 
$\mathbf{u_k}$ and the system is overdetermined (only $w+1$ variables), we could 
always find an exact solution for \eqref{eq:WENOgcondition} in our tests. 
In regions where the solution is not smooth, the weights should be chosen such that 
the smoothest polynomial of $\mathbf{p}_i$ is preferred. 
For this purpose, we use a smoothness indicator as suggested in~\cite{JiaShu96}:
\begin{align}
  \boldsymbol{\beta}_i = \sum_{l=1}^w \int_{I_j} \Delta x^{2l-1} \left( \frac{d^l}{dx^l} \mathbf{p}_i(x) \right)^2 \text{d}x.
\label{eq:WENOsmoothness}
\end{align}
Because $\boldsymbol{\beta}_i$ is large for non-smooth $\mathbf{p}_i$, the weights are chosen indirect proportional to
$\boldsymbol{\beta}_i$. 
We use either the standard WENO choice
\begin{align}
  \boldsymbol{\tilde{\omega}}_i(x) = \frac{\gamma_i(x)}{\left( 10^{-6} + \boldsymbol{\beta}_i \right)^2},
\label{eq:WENOweights}
\end{align}
or the improved WENO-Z version~\cite{BorCarCos08} for $w=2$,
\begin{align}
  \boldsymbol{\tilde{\omega}}_i(x) = \gamma_i(x) \left( 1 + \frac{|\boldsymbol{\beta_3} - \boldsymbol{\beta_1}|}{\boldsymbol{\beta}_i + 10^{-40}} \right),
\label{eq:WENOZweights}
\end{align}
and normalize the results:
\begin{align}
  \boldsymbol{\omega}_i(x) = \frac{ \boldsymbol{\tilde{\omega}}_i(x) }{\sum_{l=1}^{w+1} \boldsymbol{\tilde{\omega}}_l(x)} \quad, 
\label{eq:WENOweightsnorm}
\end{align}
where $\boldsymbol{\omega}_i(x)$ are the final reconstruction weights.
The reconstructed solution is then given by:
\begin{align}
  \mathbf{u^{WENO}}(x) = \sum_{i=1}^{w+1} \boldsymbol{\omega}_i(x) \mathbf{p}_i(x).
\label{eq:WENOapprox} 
\end{align}
 
To generalize the presented reconstruction mechanism to 2D and 3D, 
we use the procedure described in~\cite{ZhaTan13}. For simplicity, we
assume a rectilinear 2D grid structure with $N+1$ grid points $\xi^p$ per cell and direction. 
To reduce the full reconstruction of the cell $I_{jk}$ to the 1D case, 
we decouple the different directions.
First we perform $2w+1$ 1D WENO reconstructions in the $x$ direction with input data
\begin{align}
  \left\{\mathbf{u_{j-w,\tilde{k}}}, \mathbf{u_{j-w+1,\tilde{k}}}, \cdots, \mathbf{u_{j+w,\tilde{k}}}\right\}, \quad k-w \leq \tilde{k} \leq k+w
\label{eq:2DWENOx1}
\end{align}
to reconstruct the $N+1$ averages per cell:
\begin{align}
  \mathbf{u_{j,\tilde{k}}^p} := \int_{I_{j\tilde{k}}} \mathbf{u}(\xi_p,y,t) \ dy,& \quad k-w \leq \tilde{k} \leq k+w,  \label{eq:2DWENOx2}\\
                                                                                 & \quad 1 \leq p \leq m+1 \quad.\notag
\end{align}
Then, we can apply a second 1D WENO reconstruction based on the 1D averages in $y$ direction with the input data
\begin{align}
 \left\{ \mathbf{u_{j,k-w}^p}, \mathbf{u_{j,k-w+1}^p}, \cdots, \mathbf{u_{j,k+w}^p} \right\}, \quad 1 \leq p \leq m+1
\label{eq:2DWENOy1}
\end{align}
to get the 2D reconstructed values inside the cell $I_{jk}$:
\begin{align}
  \mathbf{u^{WENO}_{jk}}(\xi_p,\xi_q), \quad 1 \leq p,q \leq m+1 \quad.
\label{eq:2DWENOy2}
\end{align}
  
%%%%%%%%%%%%%%%%%%%%%%%%%%%%%%%%%%%%%%%%%%%%%%%%%%%%%%%%%%%%%%%%%%%%%%%%%%%%%%%%%%%%%%%%%%%%%%%%%%
\subsubsection{Simple WENO reconstruction}
\label{sec:simple_WENO}

To reconstruct the polynomial with a standard WENO method, the cell averages of many neighboring 
cells are needed. This leads to large computational costs and an undesirable smoothing of the solution. 
In~\cite{ZhoShu13}, it is discussed, that this standard procedure is not necessary in a DG method, since the neighboring cells yield 
more information than only a cell average value. This idea implies the simple WENO reconstruction, in which the standard
WENO methodology is applied to the cell polynomial and the neighboring cell polynomials by redefining 
\begin{align}
  \mathbf{p}_i(x) := \ &\mathbf{u}|_{I_{j+i}}(x) \ +   \label{eq:simpleWENO}\\ 
                       &\frac{1}{\Delta x}\int_{I_j} \left( \mathbf{u}|_{I_{j}}(x) - \mathbf{u}|_{I_{j+i}}(x) \right) dx, \quad i=-1,0,1. \notag
\end{align}
The integral values cause a shift of the polynomials, so that they all have the same average value in cell $I_j$ and the
cell average is conserved during reconstruction. The corresponding expressions in \eqref{eq:WENOsmoothness} and \eqref{eq:WENOapprox} 
have to be substituted. Furthermore, we only use the next neighbors, which is setting $w=1$ (3 cell stencil) in all WENO formulas. 
Since with the new ansatz \eqref{eq:simpleWENO} every linear combination of the $\mathbf{p}_i(x)$ is a higher order approximation, there is no
need to find special ideal weights as in the standard WENO method. Instead, we can freely choose the weights for all
involved cells. In our tests, we choose $\gamma_{-1} = \gamma_{1} = 1$\e{-5}, $\gamma_0 = 1- 2 \gamma_1 = 1-2$\e{-5} for smooth setups and
$\gamma_{-1} = \gamma_{1} = 1$\e{-3}, $\gamma_0 = 1- 2 \gamma_1= 1-2$\e{-3} for problems with discontinuities. 
 
%%%%%%%%%%%%%%%%%%%%%%%%%%%%%%%%%%%%%%%%%%%%%%%%%%%%%%%%%%%%%%%%%%%%%%%%%%%%%%%%%%%%%%%%%%%%%%%%
\subsection{Subcell evolution method}
\label{sec:subcell_method}
%%%%%%%%%%%%%%%%%%%%%%%%%%%%%%%%%%%%%%%%%%%%%%%%%%%%%%%%%%%%%%%%%%%%%%%%%%%%%%%%%%%%%%%%%%%%%%%%%%

Finally, we consider a hybrid FD-DG method motivated by~\cite{DumZanLou14} where shock capturing is performed on a subgrid of
equidistant grid points.  The method of~\cite{DumZanLou14} is based on
subgrids, an a posteriori troubled cell indicator, and a locally
implicit time integrator. We decided to investigate the subgrid method
separately without these other features, so we cannot compare directly
to~\cite{DumZanLou14}. There are open questions regarding the
stability and accuracy of the subgrid method, in particular when used
without the locally implicit time integration method (but see also~\cite{SonMun14,HueCasPer12}).
For our method we use the same troubled cell indicator as introduced
in~\ref{sec:WENO_method}. If a cell $I_j$ has been marked as troubled,
we subdivide this cell in $2N+1$ equidistant subcells $J_k$ containing a
single point $y_k$ each (where $N$ is the polynomial order) and compute 
the value of the approximating polynomial on the individual subcells points:
\begin{align}
  \mathbf{v_{k}} = \mathbf{\hat{u}}_{p} \phi^{p}(y_k), \qquad \text{for all } y_k \in I_j
\label{eq:subcell_averages}.
\end{align}
This map $\mathbf{\hat{u}}_{p} \mapsto \mathbf{v_{k}}$ can be done
with the subcell projection operator $\mathcal{P}$.  The back
projection $\mathcal{P}^{-1}$ is non-trivial, because the problem of
finding a polynomial of order $N$ to satisty the given $2N+1$
equations~\eqref{eq:subcell_averages} is overdetermined. 
Performing a least-squares fit of a $N$-th order polynomial for the
$2N+1$ points turns out to be a good choice for a back
projection. In our tests, we found the corresponding matrices for
$\mathcal{P}$ and $\mathcal{P}^{-1}$ to be pseudoinverse. This is easy
to verify, because whenever $\mathbf{v_{k}}$ originate
from an exact $N$-th order polynomial, a least-squares fit
$\mathcal{P}^{-1}$ will give the exact polynomial coefficients, so
$\mathcal{P}^{-1} \mathcal{P} = \mathbf{1}$ (but not neccessarily
$\mathcal{P} \mathcal{P}^{-1} = \mathbf{1}$).  
It is important to notice that contrary to~\cite{DumZanLou14}, we use a 
projection matrix based on the point values in the subcells, not 
the averages $\mathbf{v_{k}}$. This is necessary, since we want to 
employ a FD code on the subcells instead of a FV method, leading to a 
violation of conservation laws (e.g. of the rest-mass), when a projection
from topcell to subcells, or vice versa is done. However, in our tests
we found this defects decaying with order $N+1$, when we raise the grid
resolution. 
We import all necessary routines of the
BAM code~\cite{BruGonHan06,ThiBerBru11,DieBerUje15}.
In~\cite{ThiBerBru11,BerDie15} this scheme is explained in detail. 
Further improvements allow to obtain high order convergence in smooth regions will 
be presented in~\cite{BerDie15}.
The general idea is to discretize Eq.~\eqref{eq:con_law} as
\begin{align}
  \partial_t \mathbf{u_k} = \frac{2N+1}{\Delta x} \left( \mathbf{F_{k-\frac{1}{2}}} 
                      - \mathbf{F_{k+\frac{1}{2}}} \right) + \mathbf{S_k}
\label{eq:bam_scheme}
\end{align}
with $\Delta x$ being the cell grid spacing, $\frac{\Delta x}{2N+1}$ 
the subcell grid spacing and 
$\mathbf{F_{k+\frac{1}{2}}}$ the numerical flux at the boundary between
subcells $J_k$ and $J_{k+1}$. 
  
The subcell interface values of the fluxes
$\mathbf{f_{k\pm\frac{1}{2}}}$ are computed with the LLF scheme.
The necessary right and left states for the interface flux calculation
are provided by a WENOZ~\cite{BorCarCos08,JiaShu96} reconstruction
from the given subcell values.  Having evaluated the RHS of
Eq.~\eqref{eq:bam_scheme}, we use an explicit fourth-order Runge-Kutta
method for the time step.  After each Runge-Kutta substep the new
subcell values are back projected to the DG-grid by
$\mathcal{P}^{-1}$. If the cell stays troubled in the next time step,
the next evolution step is based on the subcell results without 
using the back-projected results.

%%%%%%%%%%%%%%%%%%%%%%%%%%%%%%%%%%%%%%%%%%%%%%%%%%%%%%%%%%%%%%%%%%%%%%%%%%%%%%%%%%%%%%%%%%%%%%%%%%
%%%%%%%%%%%%%%%%%%%%%%%%%%%%%%%%%%%%%%%%%%%%%%%%%%%%%%%%%%%%%%%%%%%%%%%%%%%%%%%%%%%%%%%%%%%%%%%%%%
\section{Numerical Implementation}
\label{sec:numerics}
%%%%%%%%%%%%%%%%%%%%%%%%%%%%%%%%%%%%%%%%%%%%%%%%%%%%%%%%%%%%%%%%%%%%%%%%%%%%%%%%%%%%%%%%%%%%%%%%%%
%%%%%%%%%%%%%%%%%%%%%%%%%%%%%%%%%%%%%%%%%%%%%%%%%%%%%%%%%%%%%%%%%%%%%%%%%%%%%%%%%%%%%%%%%%%%%%%%%%

Throughout this article we employ the \bamps code~\cite{HilWeyBru15}. 
It is based on the method-of-lines with a pseudospectral decomposition in the spatial part and an explicit 
fourth-order Runge-Kutta for the time stepping. 
It has been successfully used to study the gravitational wave collapse and it allows long-term simulations of 
single black hole spacetimes with excision techniques. 
The program exhibits a hybrid p-thread/MPI parallelization strategy and shows almost ideal scaling for up to 
several thousands of computing cores in vacuum simulations, 
see~\cite{HilWeyBru15} for more details. 

In this work we extend the \bamps code by implementing 
(i)~discontinuous Galerkin methods,
(ii)~a general relativistic hydrodynamics scheme for fixed background metrics, 
(iii)~a simple high resolution shock capturing (HRSC) scheme as in~\cite{ZhaTan13},
(iv)~a subcell-HRSC scheme~\cite{DumZanLou14}.
This work is the first step towards a more general infrastructure for the simulation 
of compact binary systems where matter is present. 

Although \bamps allows grid structures known as "cubed spheres"~\cite{RonIacPao96}, 
we restrict ourselves to simple Cartesian boxes. However, a generalization could be achieved easily. 
For the actual implementation of Eq.~\eqref{eq:weak_form}, we map each $I_j$ to a reference box $[-1,1]^3$ and 
define $N+1$ Legendre-Gauss-Lobatto points $\xi^p \in [-1,1]$ for each direction. Given these points, we choose
the basis $v^{k}$ of $\mathbb{P}^N([-1,1]^3)$ to be the product of the corresponding Lagrange interpolating 
polynomials each applied to one component of $\xi$
\begin{equation}
  v^k \equiv v^{pqr} = \ell^p \ell^q \ell^r
  \label{eq:basis}
\end{equation}
with
\begin{equation}
  \ell^p(\xi) = \prod_{j=1 \atop j\neq p}^{N+1} \frac{\xi-\xi^j}{\xi^p-\xi^j},
  \label{eq:LIP}
\end{equation}
i.e. we use a nodal DG formulation. 
The chosen basis allows us to use $\ell^p(\xi^r) = \delta^{pr}$ and simplifies the computation of the coefficients 
$\mathbf{\hat{u}}_{pqr}=\mathbf{u}(\xi^p,\xi^q,\xi^r)$ (interpolation condition). In contrast to the modal DG formulation, 
the flux and source coefficients are then easily determined by pointwise evaluations 
$\mathbf{\hat{f}}_{pqr}^i = \mathbf{f}^i(\mathbf{\hat{u}}_{pqr})$, 
$\mathbf{\hat{S}}_{pqr} = \mathbf{S}(\mathbf{\hat{u}}_{pqr})$.
Defining the mass matrix 
\begin{equation}
  M^{ab}=\int_{-1}^{1} \ell^a(\xi) \ell^b(\xi) d\xi
  \label{eq:massm}
\end{equation}
and the stiffness matrix 
\begin{equation}
  S^{ab}=\int_{-1}^{1} \partial_\xi \ell^a(\xi) \ell^b(\xi) d\xi
  \label{eq:stiffnm}
\end{equation}
we seperate analytic expressions and numerical variables in Eq.~\eqref{eq:weak_form} 
to gain the semidiscrete scheme:
\begin{alignat}{3}
  \partial_t & \mathbf{\hat{u}}_{pqr} = &&\notag\\
  &+\frac{2}{\Delta x}&&\left( M^{-1}_{pa} S^{ab} \mathbf{\hat{f}}_{bqr}^1 \right. \notag\\
  & &&\left. \phantom{\mathbf{\hat{f}}} - M^{-1}_{pN} \mathbf{f}^{*1}n_1(1,\xi^q,\xi^r) - M^{-1}_{p0} \mathbf{f}^{*1}n_1(-1,\xi^q,\xi^r)\right) \notag\\
  &+\frac{2}{\Delta y}&& \left( M^{-1}_{qa} S^{ab} \mathbf{\hat{f}}_{pbr}^2 \right. \notag\\
  & &&\left. \phantom{\mathbf{\hat{f}}} - M^{-1}_{qN} \mathbf{f}^{*2}n_2(\xi^p,1,\xi^r) - M^{-1}_{q0} \mathbf{f}^{*2}n_2(\xi^p,-1,\xi^r)\right) \notag\\
  &+\frac{2}{\Delta z}&& \left( M^{-1}_{ra} S^{ab} \mathbf{\hat{f}}_{pqb}^3 \right. \notag\\
  & &&\left. \phantom{\mathbf{\hat{f}}} - M^{-1}_{rN} \mathbf{f}^{*3}n_3(\xi^p,\xi^q,1) - M^{-1}_{r0} \mathbf{f}^{*3}n_3(\xi^p,\xi^q,-1) \right)\notag \\
  &+\mathbf{\hat{S}}_{pqr}.
  \label{eq:semidiscrete_DG}
\end{alignat}
Due to the choice of collocation points, the mass and stiffness matrix can be determined using
Legendre-Gauss-Lobatto integration with the corresponding weights $\omega^p$:
\begin{align}
  M^{ab}&\approx \delta^{ab} \cdot \omega^a   \label{eq:mmatrix_approx}\\
  S^{ab}&= \partial_\xi \ell^a(\xi^b) \cdot \omega^b.
  \label{eq:smatrix_approx}
\end{align}
Notice, that Eq.~\eqref{eq:mmatrix_approx} is just an approximation, while~\eqref{eq:smatrix_approx} is exact, since 
the $N+1$-point Legendre-Gauss-Lobatto integration is exact for polynomials of order $2N-1$. This approximation
simplifies the scheme and brings $M$ in a diagonal form. Furthermore, it is equal to a modal filter, which decreases 
the highest mode by a factor~$N/(2N+1)$~\cite{GasKop11}.

For the standard WENO implementation, we recast the crucial equations in matrix form, where all matrices
can be precomputed from the geometry before evolution. During the actual simulation
(i) the smoothness indicators are calculated from the cell averages as a quadratic form 
   $\beta_i = Q^{kl}_i u_{i+k} u_{i+l}$;
(ii) the weights are determined by~\eqref{eq:WENOweightsnorm}; 
(iii) the value $p_i(\xi^q)$ of the approximating polynomial of stencil $i$ at the collocation points
$\xi^q$ is evaluated from the cell averages by a matrix-vector multiplication 
$p_i(\xi^q) = C^{qr}_i u_{i+r}$ originating from \eqref{eq:WENOpcondition};
(iv) the final reconstruction is calculated by \eqref{eq:WENOapprox}.
For simple WENO computations, the only difference is that in steps (i) and (iii) the matrices are
larger, because the $\beta_i$ and $p_i(\xi^q)$ do not only depend on the averages of the neighbor
cells, but the full polynomial given by $N+1$ coefficients per cell.

In contrast to previous work, where no restriction algorithm (Sec.~\ref{sec:WENO_method})
was present, we need to communicate more then just the two-dimensional boundary layers of every cell. 
Therefore, we introduced a new grid distribution method to reduce the communication between different processors. 
The Cartesian grid consisting of $n=n_x n_y n_z$ boxes is distributed on $p$ processes 
in a way that communication between the processes is minimal.
For this purpose, we perform a prime decomposition of $p = p_1 p_2 ... p_i$ and set the number of grids per
direction $p_x = p_1, p_y = p_2, p_z = p_3$ initially. Let $p_\text{min} = \min(p_x,p_y,p_z)$, we recalculate $p_\text{min}$
as $p_\text{min} \mapsto p_\text{min} \cdot p_4$. For further $p_j$, we proceed in the same manner, so that each $p_j$ is
always multiplied with the smallest of $p_x, p_y, p_z$.
Finally, we subdivide the full box grid in $p_x$ parts in $x$ direction, 
in $p_y$ parts in $y$ direction and in $p_z$ parts in $z$ direction.
Each of these $p_x \cdot p_y \cdot p_z$ parts is mapped to one MPI-process, 
which gives a simple box decomposition, which is almost cubical.

%%%%%%%%%%%%%%%%%%%%%%%%%%%%%%%%%%%%%%%%%%%%%%%%%%%%%%%%%%%%%%%%%%%%%%%%%%%%%%%%%%%%%%%%%%%%%%%%%%
%%%%%%%%%%%%%%%%%%%%%%%%%%%%%%%%%%%%%%%%%%%%%%%%%%%%%%%%%%%%%%%%%%%%%%%%%%%%%%%%%%%%%%%%%%%%%%%%%%
\section{Simple testbeds}
\label{sec:simple_tests}
%%%%%%%%%%%%%%%%%%%%%%%%%%%%%%%%%%%%%%%%%%%%%%%%%%%%%%%%%%%%%%%%%%%%%%%%%%%%%%%%%%%%%%%%%%%%%%%%%%
%%%%%%%%%%%%%%%%%%%%%%%%%%%%%%%%%%%%%%%%%%%%%%%%%%%%%%%%%%%%%%%%%%%%%%%%%%%%%%%%%%%%%%%%%%%%%%%%%%

\begin{table}[t]
    \setlength{\tabcolsep}{1pt}
  \centering
  \caption{Numerical errors and convergence orders for the advection equation problem \eqref{eq:Advection1} at $t=10$
  for different numbers of grid patches $n_x$ and orders of DG polynomials $N$ (CFL$=0.25$, $M=1$).}
  \begin{tabular}{ l l| l l l l l l l l l}
   \hline
       &   & & \multicolumn{2}{l}{DG} & & \multicolumn{2}{l}{DG+WENO-7} & & \multicolumn{2}{l}{DG+simple WENO} 
           \\ \cline{4-5}\cline{7-8}\cline{10-11}
    $n_x$ & $N$ & & $L_1$ error  & order
           & & $L_1$ error \qquad & order & & $L_1$ error \qquad & order \\ \hline
    10  & 1 & & $1.53$\e{-1 } & -    & & $2.92$\e{-1 } & -    & & $1.51$\e{-1 } & -   \\
    20  &   & & $3.63$\e{-2 } & 2.08 & & $1.40$\e{-1 } & 1.05 & & $4.79$\e{-2 } & 1.65 \\
    40  &   & & $5.34$\e{-3 } & 2.76 & & $3.77$\e{-2 } & 1.90 & & $6.64$\e{-3 } & 2.85 \\
    80  &   & & $7.29$\e{-4 } & 2.87 & & $7.15$\e{-3 } & 2.39 & & $7.29$\e{-4 } & 3.18 \\
    160 &   & & $1.24$\e{-4 } & 2.54 & & $1.28$\e{-3 } & 2.48 & & $1.24$\e{-4 } & 2.54 \\
    320 &   & & $2.84$\e{-5 } & 2.12 & & $2.36$\e{-4 } & 2.43 & & $2.84$\e{-5 } & 2.12 \\
    \hline
    10  & 3 & & $2.04$\e{-4 } & -    & & $4.84$\e{-2 } & -    & & $2.12$\e{-4 } & -   \\
    20  &   & & $1.02$\e{-5 } & 4.32 & & $1.51$\e{-3 } & 4.99 & & $1.02$\e{-5 } & 4.37 \\
    40  &   & & $6.27$\e{-7 } & 4.03 & & $4.99$\e{-5 } & 4.92 & & $6.36$\e{-7 } & 4.01 \\
    80  &   & & $3.90$\e{-8 } & 4.00 & & $9.71$\e{-7 } & 5.68 & & $3.98$\e{-8 } & 3.99 \\ 
    160 &   & & $2.44$\e{-9 } & 4.00 & & $1.60$\e{-8 } & 5.92 & & $2.53$\e{-9 } & 3.97 \\
    320 &   & & $1.52$\e{-10} & 4.00 & & $3.03$\e{-10} & 5.72 & & $1.62$\e{-10} & 3.96 \\
    \hline
    10  & 5 & & $7.99$\e{-7 } & -    & & $9.95$\e{-2 } & -    & & $5.75$\e{-6 } & -   \\
    20  &   & & $1.88$\e{-8 } & 5.40 & & $1.11$\e{-2 } & 3.15 & & $1.27$\e{-7 } & 5.49 \\
    40  &   & & $8.92$\e{-10} & 4.39 & & $4.09$\e{-4 } & 4.76 & & $5.26$\e{-9 } & 4.59 \\
    80  &   & & $5.42$\e{-11} & 4.04 & & $8.67$\e{-6 } & 5.56 & & $1.28$\e{-10} & 5.35 \\
    160 &   & & $3.40$\e{-12} & 3.99 & & $1.40$\e{-7 } & 5.95 & & $1.52$\e{-11} & 3.07 \\
    320 &   & & $6.48$\e{-13} & 2.39 & & $7.66$\e{-10} & 7.51 & & $1.94$\e{-10} & -    \\
    \hline
  \end{tabular}
  \label{tab:Advection1}
\end{table}

\subsection{Advection equation}
\label{sec:advection_equation}
As a first test for our new algorithms we consider the advection equation
without a source ($S=0$)
\begin{align}
 \partial_t u + \partial_x u = 0
 \label{eq:Advection}
\end{align}
for a Gaussian peak on the interval $x\in[-1,1]$ 
\begin{align}
  \psi(x,0) = 
  A\mathrm{e}^{(-x^2 / \sigma^2)} + 
  A\mathrm{e}^{(-(x-2)^2 / \sigma^2)} +
  A\mathrm{e}^{(-(x+2)^2 / \sigma^2)}
\label{eq:Advection1}
\end{align}
(we artificially add two peaks to gain smooth, periodic initial data) and a 
rectangular pulse (non-smooth initial data)
\begin{align}
 \psi(x,0) = 
 \begin{cases}
   1 & \text{if } |x - x_0| < \sigma \\
   0 & \text{else }
 \end{cases} \quad .
\label{eq:Advection2}
\end{align}
The convergence rate in the first test case ($A=1$, $\sigma=0.4$) is influenced by several effects; see Table.~\ref{tab:Advection1}. 
 
For our choice of polynomials with order $N$, we find convergence rates up to order $N+1$, as expected. 
However, in an error regime beyond $10^{-10}$, we observe a further drop in the convergence rates, 
because of the growing influence of truncation errors. Applying the standard WENO reconstruction
procedure leads to slightly different results. Convergence for small numbers of $n_x$ is slower, 
but finally shows convergence above $N+1$-th order. 
This can be explained by the decreasing influence of the WENO procedure for increasing $n_x$. 
The cell indicator only marks the cells around the maximum of the Gaussian peak as troubled, 
so the effective area, where the WENO reconstruction takes place, decreases. Since the reconstruction has
a strong smoothing effect, the numerical results significantly differ from the analytic solution 
for small $n_x$ and tend to the pure DG solution for large $n_x$. 
Comparing the two WENO-implementations, we observe that the simple WENO algorithm shows slower convergence, but 
while the standard WENO is $\sim1-2$ orders of magnitude less accurate than the pure DG-evolution, 
the simple WENO performs much better, showing roughly the same $L_1$-errors as the pure DG-evolution.

For the second case Eq.~\eqref{eq:Advection2}, which we just want to summarize briefly, 
we observe larger total errors than for the smooth problem discussed above. 
Again the pure DG-method errors are below the corresponding errors for the DG + standard WENO method. 
However, the difference are at most a factor of $2$. The simple WENO algorithm has comparable errors as the 
DG + standard WENO method. Independent of the scheme we observe first order convergence, which 
is consistent with the expectation for a  non-smooth problem containing discontinuities. 
  
\begin{figure}[t]
  \centering
  \includegraphics[width=0.5\textwidth]{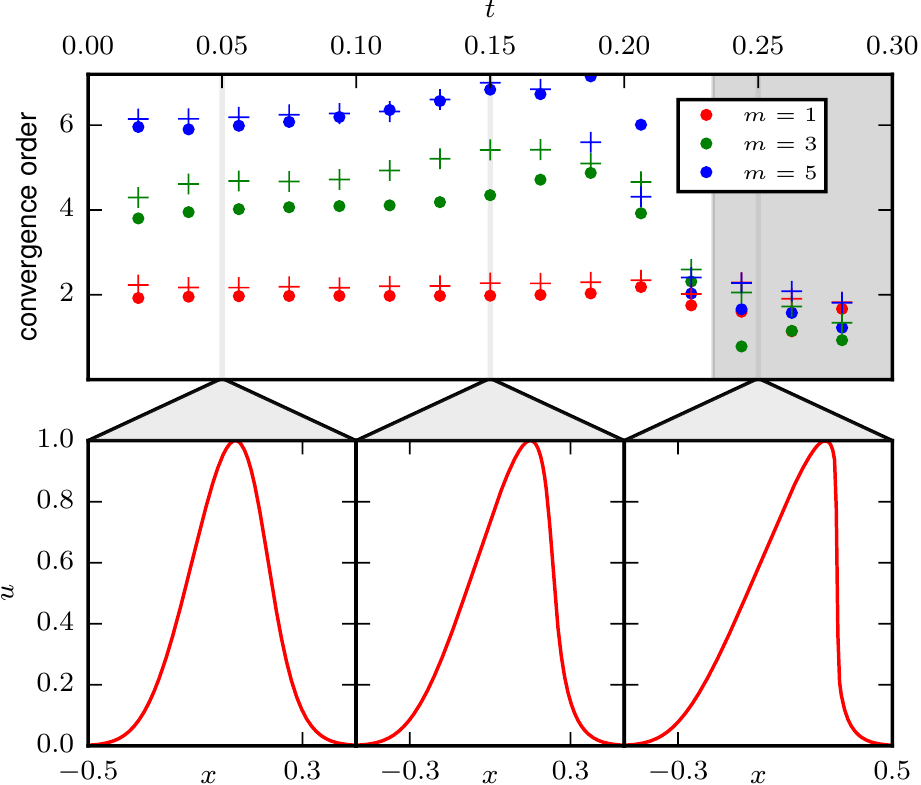}
  \caption{Convergence rate during the evolution of a Gaussian wave packet for the Burgers equation \eqref{eq:Burgers}:
  As expected, the convergence rate is around $N+1$ during the evolution of a smooth wave. At $t_\text{shock} \approx 0.233$, a shock
  forms and the rates significantly drop down to first order convergence. When a standard WENO-7 reconstruction is used (crosses)
  the convergence rates are slightly higher than for the pure DG scheme (dots). The convergence rate is calculated from the errors of
  a $n_x = 160$ and a $n_x = 320$ run.}
  \label{fig:Burgers}
\end{figure}

%%%%%%%%%%%%%%%%%%%%%%%%%%%%%%%%%%%%%%%%%%%%%%%%%%%%%%%%%%%%%%%%%%%%%%%%%%%%%%%%%%%%%%%%%%%%%%%%%%
\subsection{Burgers equation}
\label{sec:burger_equation}
%%%%%%%%%%%%%%%%%%%%%%%%%%%%%%%%%%%%%%%%%%%%%%%%%%%%%%%%%%%%%%%%%%%%%%%%%%%%%%%%%%%%%%%%%%%%%%%%%%

The Burgers equation without source ($S=0$)
\begin{align}
 \partial_t u + u \partial_x u = 0
 \label{eq:Burgers}
\end{align}
allows the formation of shocks from smooth initial data $u_0$. 
After the time
\begin{align}
  t_\text{shock} = -\left( \min \frac{\partial u_0}{\partial x} \right)^{-1}
\end{align}
shocks will appear during the evolution. 
We use this as a testbed for our code and evolve the initial Gaussian 
peak~\eqref{eq:Advection1} with $A=1$ and $\sigma=0.2$. 
For this initial conditions, a shock forms at $t_\text{shock} \approx 0.23316$. 
Similarly to our results for the advection equation 
we observe that the convergence rate is decreasing after $t_{\rm shock}$; cmp.~Fig.~\ref{fig:Burgers}. 
The upper panels show snapshots of the field $u$ at times $t=0.05,0.15,0.25$.
Without WENO-reconstruction (circles) 
we observe the expected convergence order of $N+1$ up to $t_{\rm shock}$. 
Shortly before the shock formation at $t_{\rm shock}$ convergence start to drop for all $N$ (gray shaded region). 
Employing a standard WENO algorithm convergence is slightly above 
the expected $N+1$-th order. As discussed for the advection equation, this is related to 
the amount of troubled cells, which are reconstructed. 
For higher resolution a smaller percentage of cells is reconstructed and 
consequently a faster convergence is observed. After the shock formation the convergence 
order drops also for DG+WENO to approximately first order convergence. 
  
In addition, we prepared the initial conditions
\begin{align}
 \psi(x,0) = 0.5 + \sin(x\pi)
 \label{eq:Burgers2}
\end{align}
and check convergence at $t=0.5/\pi$ to compare with the results of~\cite{QiuShu05}, 
Tab.~\ref{tab:Burgers2} summarizes the results. Because of the smoothness of the solution, 
we observe for $N=1$ polynomials second order convergence independent of the reconstruction method applied 
in the troubled cells. While the total $L_1$-error for DG+WENO-5 is approximately a factor of 
2-3 larger than the pure DG evolution, we see that the DG+simple WENO algorithm performs as good as pure DG. 
For $N=3$ polynomials, we expect fourth order convergence, which we can verify with the pure DG and the DG+simple WENO
setup. The DG+WENO-5 algorithm shows a higher convergence rate for low resolutions, which is again caused by the fact that 
larger number of cells decrease the interval where a reconstruction is performed. 
  
\begin{table}[t]
  \centering
      \setlength{\tabcolsep}{1pt}
    \caption{Numerical errors and convergence orders for the Burgers equation problem \eqref{eq:Burgers2} at $t=\frac{0.5}{\pi}$
  for different numbers of grid patches $n_x$ and orders of DG polynomials $N$.}
  \begin{tabular}{ l l| l l l l l l l l l}
   \hline
       &   & & \multicolumn{2}{l}{DG} & & \multicolumn{2}{l}{DG + WENO-7} & & \multicolumn{2}{l}{DG + simple WENO} 
           \\ \cline{4-5}\cline{7-8}\cline{10-11}
    $n_x$&$N$& & $L_1$ error & order
           & & $L_1$ error& order & & $L_1$ error & order \\ \hline
    10  & 1 & & $5.34$\e{- 2} & -    & & $6.09$\e{- 2} & -    & & $8.23$\e{- 2} & - \\
    20  &   & & $1.45$\e{- 2} & 1.87 & & $1.80$\e{- 2} & 1.75 & & $1.61$\e{- 2} & 2.34 \\
    40  &   & & $4.29$\e{- 3} & 1.76 & & $4.66$\e{- 3} & 1.94 & & $4.80$\e{- 3} & 1.75 \\
    80  &   & & $1.24$\e{- 3} & 1.78 & & $1.29$\e{- 3} & 1.84 & & $1.24$\e{- 3} & 1.94 \\
    160 &   & & $3.60$\e{- 4} & 1.78 & & $3.69$\e{- 4} & 1.80 & & $3.60$\e{- 4} & 1.78 \\
    320 &   & & $1.02$\e{- 4} & 1.82 & & $1.03$\e{- 4} & 1.83 & & $1.02$\e{- 4} & 1.82 \\
    \hline
    10  & 3 & & $1.80$\e{- 3} &      & & $3.94$\e{- 3} & -    & & $1.80$\e{- 3} & - \\
    20  &   & & $9.80$\e{- 5} & 4.20 & & $1.50$\e{- 4} & 4.71 & & $9.80$\e{- 5} & 4.20 \\
    40  &   & & $6.36$\e{- 6} & 3.94 & & $6.72$\e{- 6} & 4.48 & & $6.36$\e{- 6} & 3.94 \\
    80  &   & & $4.21$\e{- 7} & 3.91 & & $4.22$\e{- 7} & 3.99 & & $4.21$\e{- 7} & 3.91 \\
    160 &   & & $2.71$\e{- 8} & 3.95 & & $2.71$\e{- 8} & 3.95 & & $2.71$\e{- 8} & 3.95 \\
    320 &   & & $1.75$\e{- 9} & 3.95 & & $1.75$\e{- 9} & 3.95 & & $1.75$\e{- 9} & 3.95 \\
    \hline
    10  & 5 & & $3.58$\e{- 5} & -    & & $5.86$\e{- 3} & -    & & $3.58$\e{- 5} & - \\
    20  &   & & $7.61$\e{- 7} & 5.55 & & $1.49$\e{- 4} & 5.29 & & $7.60$\e{- 7} & 5.55 \\
    40  &   & & $1.61$\e{- 8} & 5.56 & & $1.33$\e{- 6} & 6.81 & & $1.62$\e{- 8} & 5.54 \\
    80  &   & & $2.97$\e{-10} & 5.75 & & $1.16$\e{- 8} & 6.83 & & $2.98$\e{-10} & 5.77 \\
    160 &   & & $5.47$\e{-12} & 5.76 & & $1.63$\e{-10} & 6.14 & & $5.47$\e{-12} & 5.76 \\
    320 &   & & $1.15$\e{-13} & 5.56 & & $9.49$\e{-13} & 7.43 & & $1.15$\e{-13} & 5.57 \\    
    \hline
  \end{tabular}
  \label{tab:Burgers2}
\end{table}

%%%%%%%%%%%%%%%%%%%%%%%%%%%%%%%%%%%%%%%%%%%%%%%%%%%%%%%%%%%%%%%%%%%%%%%%%%%%%%%%%%%%%%%%%%%%%%%%%%
%%%%%%%%%%%%%%%%%%%%%%%%%%%%%%%%%%%%%%%%%%%%%%%%%%%%%%%%%%%%%%%%%%%%%%%%%%%%%%%%%%%%%%%%%%%%%%%%%%
\section{Special relativistic hydrodynamics}
\label{sec:SRHD}
%%%%%%%%%%%%%%%%%%%%%%%%%%%%%%%%%%%%%%%%%%%%%%%%%%%%%%%%%%%%%%%%%%%%%%%%%%%%%%%%%%%%%%%%%%%%%%%%%%
%%%%%%%%%%%%%%%%%%%%%%%%%%%%%%%%%%%%%%%%%%%%%%%%%%%%%%%%%%%%%%%%%%%%%%%%%%%%%%%%%%%%%%%%%%%%%%%%%%

In the following section, 
we solve the GRHD conservation law \eqref{eq:con_law}~\cite{Fon07,ThiBerBru11} without source terms and 
with $\alpha = \psi^4 = 1$
to consider special relativistic test cases, i.e.~flat spacetimes. 

%%%%%%%%%%%%%%%%%%%%%%%%%%%%%%%%%%%%%%%%%%%%%%%%%%%%%%%%%%%%%%%%%%%%%%%%%%%%%%%%%%%%%%%%%%%%%%%%%%
\subsection{One-dimensional problems}
%%%%%%%%%%%%%%%%%%%%%%%%%%%%%%%%%%%%%%%%%%%%%%%%%%%%%%%%%%%%%%%%%%%%%%%%%%%%%%%%%%%%%%%%%%%%%%%%%%
\begin{table*}[t]
  \centering
  \begin{tabular}{ l l |p{0.05cm} l l p{0.05cm} l l p{0.05cm} l l p{0.05cm} l l p{0.05cm} l l p{0.05cm} l l }
   \hline
       &   & & \multicolumn{2}{l}{DG} & & \multicolumn{2}{l}{DG + WENO-5} & & \multicolumn{2}{l}{DG + WENO-Z} & & \multicolumn{2}{l}{DG + simple WENO} 
           & & \multicolumn{2}{l}{DG + subcells} & & \multicolumn{2}{l}{subcells only}\\ \cline{4-5}\cline{7-8}\cline{10-11}\cline{13-14}\cline{16-17}\cline{19-20}
    $n_x$ & $N$ & & $L_1$ error & order 
           & & $L_1$ error & order   
           & & $L_1$ error & order
           & & $L_1$ error & order
           & & $L_1$ error & order
           & & $L_1$ error & order\\ \hline
    10  & 1 & & $1.22$\e{-3 } & -    & & $8.46$\e{-2 } & -    & & $8.62$\e{-2 } & -    & & $1.80$\e{-2 } & -    & & $2.85$\e{-3 } & -    & & $3.22$\e{-6 } & -\\
    20  &   & & $2.73$\e{-4 } & 2.15 & & $2.80$\e{-2 } & 1.59 & & $2.69$\e{-2 } & 1.67 & & $1.86$\e{-3 } & 3.27 & & $1.95$\e{-3 } & 0.54 & & $1.00$\e{-7 } & 5.00\\
    40  &   & & $6.72$\e{-5 } & 2.02 & & $4.83$\e{-3 } & 2.53 & & $4.77$\e{-3 } & 2.49 & & $7.03$\e{-5 } & 4.72 & & $4.07$\e{-4 } & 2.26 & & $3.14$\e{-9 } & 4.99\\
    80  &   & & $1.67$\e{-5 } & 2.00 & & $6.61$\e{-4 } & 2.86 & & $6.43$\e{-4 } & 2.89 & & $1.67$\e{-5 } & 2.06 & & $8.86$\e{-5 } & 2.20 & & $9.84$\e{-11} & 4.99\\
    160 &   & & $4.18$\e{-6 } & 2.00 & & $9.64$\e{-5 } & 2.77 & & $8.73$\e{-5 } & 2.88 & & $4.18$\e{-6 } & 2.00 & & $2.02$\e{-5 } & 2.13 & & $3.08$\e{-12} & 4.99\\
    320 &   & & $1.04$\e{-6 } & 2.00 & & $1.40$\e{-5 } & 2.77 & & $1.44$\e{-5 } & 2.59 & & $1.04$\e{-6 } & 2.00 & & $4.31$\e{-6 } & 2.22 & & $1.14$\e{-13} & 4.75\\\hline
    10  & 3 & & $4.27$\e{-6 } & -    & & $3.69$\e{-3 } & -    & & $9.67$\e{-4 } & -    & & $4.33$\e{-6 } & -    & & $1.58$\e{-5 } & -    & & $1.08$\e{-7 } & -\\
    20  &   & & $3.29$\e{-7 } & 3.70 & & $4.52$\e{-5 } & 6.35 & & $1.83$\e{-5 } & 5.72 & & $3.21$\e{-7 } & 3.75 & & $9.33$\e{-7 } & 4.08 & & $3.39$\e{-9 } & 4.99\\
    40  &   & & $1.79$\e{-8 } & 4.20 & & $7.37$\e{-7 } & 5.93 & & $2.10$\e{-7 } & 6.44 & & $1.76$\e{-8 } & 4.18 & & $4.49$\e{-8 } & 4.37 & & $1.06$\e{-10} & 4.99\\
    80  &   & & $9.39$\e{-10} & 4.25 & & $1.09$\e{-8 } & 6.07 & & $3.39$\e{-9 } & 5.95 & & $9.50$\e{-10} & 4.21 & & $3.56$\e{-9 } & 3.65 & & $3.31$\e{-12} & 5.00\\
    160 &   & & $6.01$\e{-11} & 3.96 & & $1.58$\e{-10} & 6.10 & & $1.31$\e{-10} & 4.69 & & $6.06$\e{-11} & 3.96 & & $2.08$\e{-10} & 4.09 & & $1.07$\e{-13} & 4.94\\
    320 &   & & $3.80$\e{-12} & 3.97 & & $6.96$\e{-12} & 4.51 & & $6.93$\e{-12} & 4.24 & & $3.84$\e{-12} & 3.97 & & $1.26$\e{-11} & 4.04 & & $2.06$\e{-14} & 2.37\\\hline
    10  & 5 & & $2.63$\e{-9 } & -    & & $8.53$\e{-3 } & -    & & $1.48$\e{-3 } & -    & & $3.79$\e{-8 } & -    & & $5.09$\e{-8 } & -    & & $1.78$\e{-8 } & -\\
    20  &   & & $3.86$\e{-11} & 6.08 & & $2.67$\e{-4 } & 4.99 & & $2.13$\e{-5 } & 6.11 & & $5.55$\e{-10} & 6.09 & & $1.36$\e{-9 } & 5.22 & & $5.57$\e{-10} & 4.99\\
    40  &   & & $6.13$\e{-13} & 5.97 & & $4.80$\e{-6 } & 5.79 & & $1.89$\e{-7 } & 6.81 & & $6.97$\e{-12} & 6.31 & & $1.04$\e{-11} & 7.02 & & $1.74$\e{-11} & 4.99\\
    80  &   & & $4.64$\e{-14} & 3.72 & & $6.47$\e{-8 } & 6.21 & & $1.54$\e{-9 } & 6.93 & & $1.95$\e{-13} & 5.15 & & $3.93$\e{-13} & 4.73 & & $5.52$\e{-13} & 4.97\\
    160 &   & & $8.63$\e{-14} & -    & & $2.98$\e{-10} & 7.76 & & $1.30$\e{-11} & 6.89 & & $9.70$\e{-14} & 1.00 & & $1.20$\e{-13} & 1.70 & & $4.41$\e{-14} & 3.64\\
    320 &   & & $1.80$\e{-13} & -    & & $8.81$\e{-13} & 8.40 & & $6.47$\e{-13} & 4.32 & & $7.69$\e{-13} & -    & & $2.48$\e{-13} & -    & & $4.94$\e{-14} & -\\
    \hline
  \end{tabular}
  \caption{Numerical errors and convergence orders for problem \eqref{eq:SineWave} at $t=2$
  for different numbers of grid patches $n_x$, orders of DG polynomials $N$ and several shock resolution methods. } 
  \label{tab:SineWave}
\end{table*}

As a first test, we consider a smooth sine wave propagating with constant speed. 
The initial conditions are:
\begin{align}
 \rho(x,t) =& 1 + 0.2 \sin (2 \pi (x-v_x t)) \notag \\
 v_x(x,t)  =& 0.2  \label{eq:SineWave} \\
 p(x,t)    =& 1 \notag 
\end{align}
inside the periodic 1D domain $x \in [-1,1]$ divided into $n_x$ uniform grid patches. 
Viewing the $L_1$ errors and convergence rates (Tab.~\ref{tab:SineWave}), we find the convergence 
rate of the DG scheme to be $N+1$, when we use polynomials $p \in \mathbb{P}^N([-1,1])$. 

Although we are dealing with a smooth problem a few cells around the maximum 
of the density $\rho$ are marked as troubled. 
When we employ the standard WENO-5 or WENO-Z reconstruction method, we observe 
at least one order of magnitude larger absolute errors as in the the pure DG-case for 
the employed resolutions.
Contrary, the convergence order is artificially higher than for the pure DG-method. 
For the simple WENO method, we obtain absolute errors compatible or identical 
with the scheme without reconstruction and obtain a convergence order of $N+1$ for an $N$-th order 
polynomial.
In case of the subcell evolution, i.e.~when we project the grid patch data on a finer
subcell treating this with finite differencing method, we observe similar convergence rates. 
 The subcell evolution itself is performed with a fifth order accurate scheme~\cite{Mig10}, 
which we verify with simulations using only subcells. 
\begin{figure}[t]
  \centering
  \includegraphics[width=0.49\textwidth]{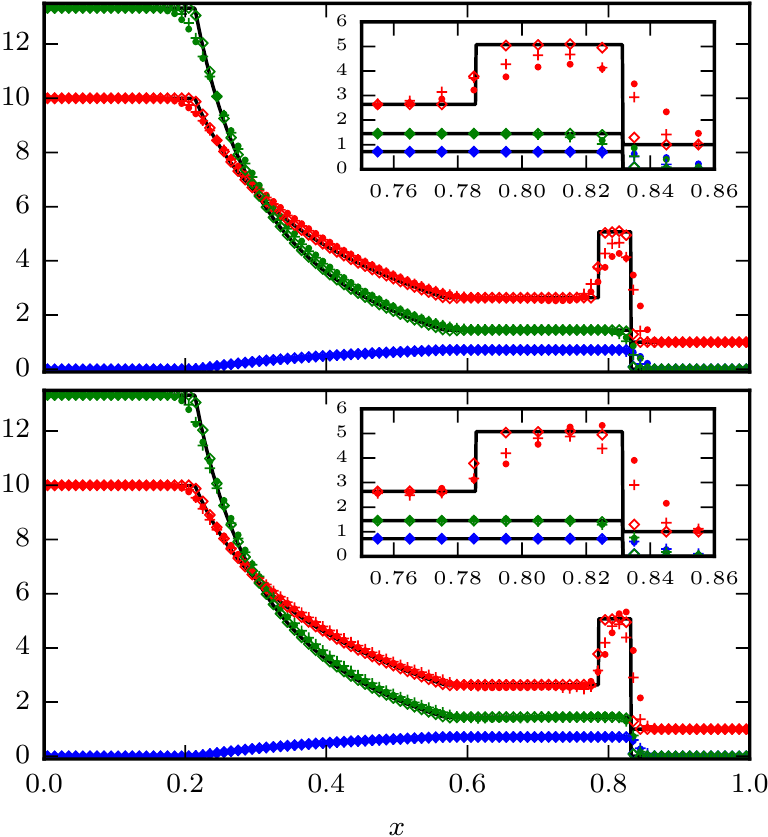}\\
  \caption{Evolution of the special relativistic shocktube initial data \eqref{eq:ShockTube} (density: red, velocity: blue, pressure: green):
  Numerical results using the standard WENO-3 (top, dots), standard WENO-5 (top, crosses), WENO-Z (bottom, dots), simple WENO (bottom, crosses) and 
  the subcell evolution method (diamonds), compared to the analytical result (black line) at $t=0.4$. For the troubled
  cell indication, we set $M=5$. }
  \label{fig:ShockTube1D}
\end{figure}

As a second test focusing on the ability of our scheme to deal with discontinuities, we consider the
shock tube problem with initial conditions
\begin{align}
  (\rho,v_x,p)(x,0) = 
  \begin{cases}
    (10,0,13.33) & \text{if } x < 0.5 \\
    (1,0,10^{-7}) & \text{if } x \geq 0.5
  \end{cases}
\label{eq:ShockTube}
\end{align}
on the domain $x \in [0,1]$. The analytical solution for this problem in the context of SRHD is given 
by~\cite{MarMul94}. During our tests, we observe the troubled cell indicator to work reliable, since the
grid patches which evolve the shock and the rarefraction wave are marked as troubled. All methods, the
standard DG-WENO methods, the simple WENO approach as well as the subcell projection method, 
are able to provide a stable evolution of 
the shock tube problem, shown in Fig.~\ref{fig:ShockTube1D}. 

%%%%%%%%%%%%%%%%%%%%%%%%%%%%%%%%%%%%%%%%%%%%%%%%%%%%%%%%%%%%%%%%%%%%%%%%%%%%%%%%%%%%%%%%%%%%%%%%%%
\subsection{Two-dimensional problems}
%%%%%%%%%%%%%%%%%%%%%%%%%%%%%%%%%%%%%%%%%%%%%%%%%%%%%%%%%%%%%%%%%%%%%%%%%%%%%%%%%%%%%%%%%%%%%%%%%%

\begin{figure*}[t]
  \centering
  \includegraphics[width=\textwidth]{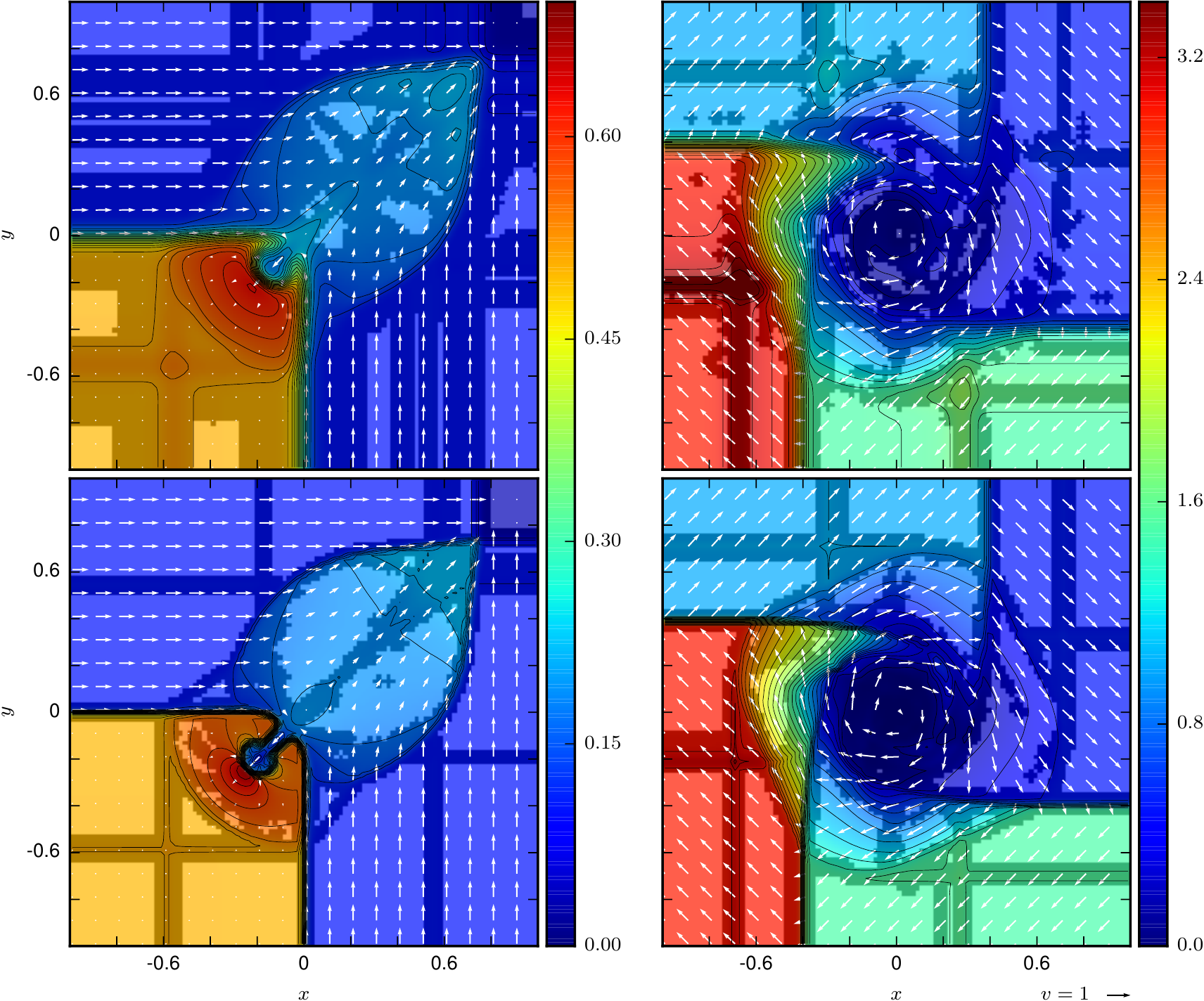}
  \caption{Special relativistic hydrodynamics simulations in 2D for the shocktube problem \eqref{eq:2DShockTube} (left) and 
  the vortex problem \eqref{eq:2DVortex} (right) at $t=0.8$, each evolved with the standard WENO-5 reconstruction (top) and
  the subcell evolution method (bottom) using $n=100 \times 100$ grid patches, polynomials of order $N=3$, $M=5$, CFL$=0.25$.
  Density plot with contours, corresponding velocity field (arrows) and troubled cells (shaded regions).}
  \label{fig:2Dsrhd}
\end{figure*}

Generalizing our results to more complex two-dimensional wave setups,
we perform two tests as presented~\cite{ZhaTan13}: 
A shock-like test with the initial conditions
\begin{align}
  (\rho,v_x,v_y,p)(x,0) = 
  \begin{cases}
    (0.03515,0,0,0.163)  & \text{if } x > 0, y > 0 \\
    (0.1,0.7,0,1)        & \text{if } x < 0, y > 0 \\
    (0.5,0,0,1)          & \text{if } x < 0, y < 0 \\
    (0.1,0,0.7,1)        & \text{if } x > 0, y < 0
  \end{cases} \quad .
\label{eq:2DShockTube}
\end{align}
and a vortex-like test with the initial conditions
\begin{align}
  (\rho,v_x,v_y,p)(x,0) = 
  \begin{cases}
    (0.5,0.5,-0.5,5.0)   & \text{if } x > 0, y > 0 \\
    (1,0.5,0.5,5.0)      & \text{if } x < 0, y > 0 \\
    (3.0,-0.5,0.5,5.0)   & \text{if } x < 0, y < 0 \\
    (1.5,-0.5,-0.5,5.0)  & \text{if } x > 0, y < 0
  \end{cases} ,
\label{eq:2DVortex}
\end{align}
with $(x,y) \in [-1,1] \times [-1,1]$. 
During the evolution of both cases, all initial discontinuities are 
captured by the troubled cell indicator. We get the results as shown 
in Fig.~\ref{fig:2Dsrhd}. 
We tested in detail the standard WENO and the DG+subcell scheme, the figures show that 
the WENO-5 and DG+subcell evolution give qualitatively the same results. 
In case of the shocktube, Eq.~\eqref{eq:2DShockTube}-- left panels, less cells are
marked troubled for the DG+subcell scheme. Furthermore, the DG+subcell 
method resolves steep gradients better than the standard WENO reconstruction. 
This becomes most dominant in a domain around $x=y=-0.2$.
However, due to the larger computational expenses the DG+subcell scheme is 
a factor of $\sim 2.4$ times slower than the standard WENO method. 
  
The right panels of Fig.~\ref{fig:2Dsrhd} represent the vortex test, cmp.~\eqref{eq:2DVortex}. 
As for the shocktube, both methods are able to resolve the structure properly.
Again the DG+subcell method gives more accurate results, i.e.~acting less dissipative 
keeping shock regions resolved, but also needs more computational resources and is $\sim 3.2$
times slower than the standard WENO implementation. 
  
%%%%%%%%%%%%%%%%%%%%%%%%%%%%%%%%%%%%%%%%%%%%%%%%%%%%%%%%%%%%%%%%%%%%%%%%%%%%%%%%%%%%%%%%%%%%%%%%%%
%%%%%%%%%%%%%%%%%%%%%%%%%%%%%%%%%%%%%%%%%%%%%%%%%%%%%%%%%%%%%%%%%%%%%%%%%%%%%%%%%%%%%%%%%%%%%%%%%%
\section{General relativistic hydrodynamcis}
\label{sec:GRHDC}
%%%%%%%%%%%%%%%%%%%%%%%%%%%%%%%%%%%%%%%%%%%%%%%%%%%%%%%%%%%%%%%%%%%%%%%%%%%%%%%%%%%%%%%%%%%%%%%%%%
%%%%%%%%%%%%%%%%%%%%%%%%%%%%%%%%%%%%%%%%%%%%%%%%%%%%%%%%%%%%%%%%%%%%%%%%%%%%%%%%%%%%%%%%%%%%%%%%%%

\begin{figure*}[t]
  \centering
  \includegraphics[width=\textwidth]{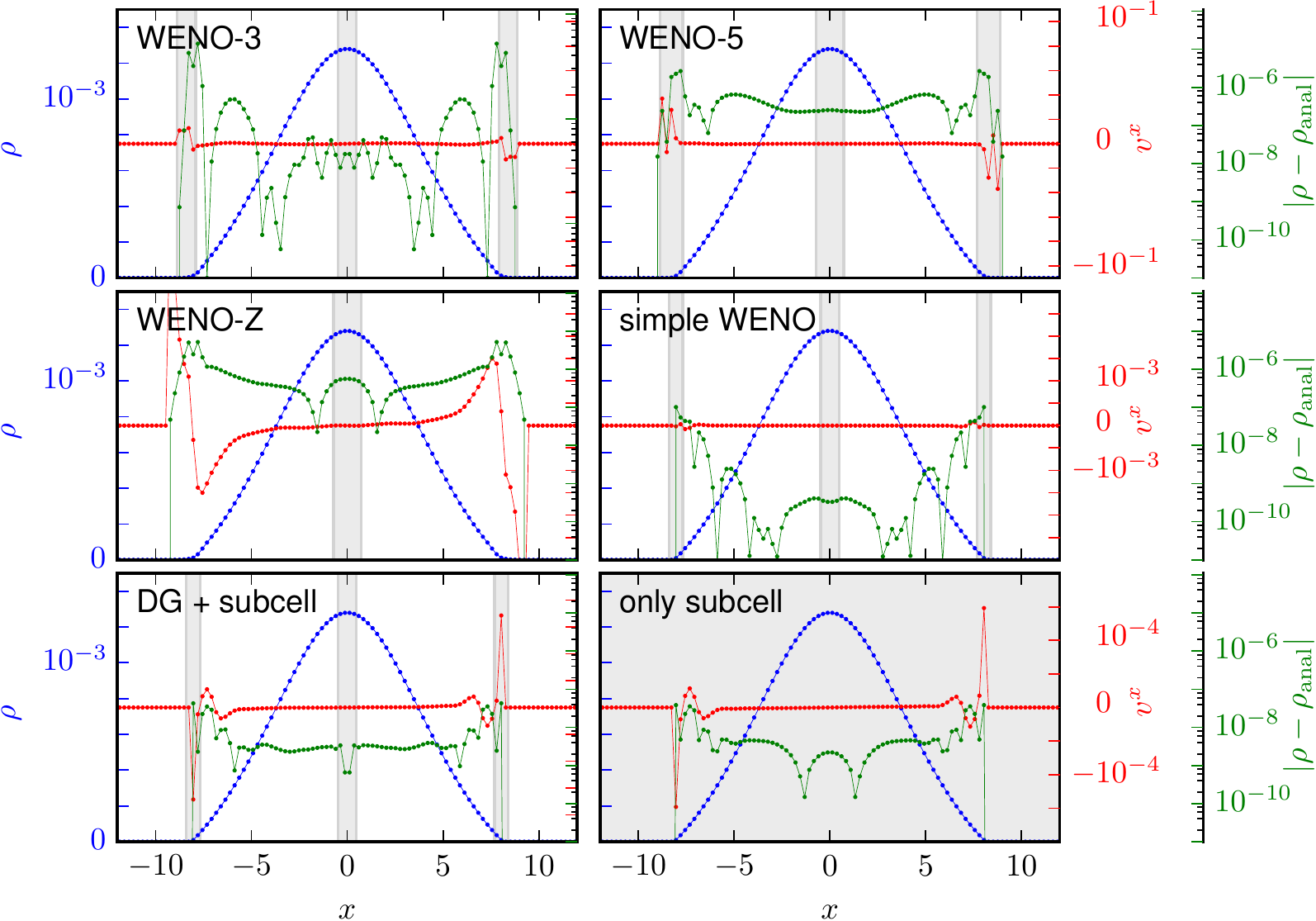}
  \caption{Density (blue), density error (green) and velocity (red) cell averages at $t=1000$ for a 1D TOV star using $n=100$ cells, polynomials of order $N=3$,
  CFL $= 0.25$, $f_{atm} = 1$\e{-8} and $f_{thr} = 100$ evolved with DG and several shock resolution methods. The cells marked as troubled at $t=1000$ are
  colored in gray. The result of the pure subcell run, which is equivalent to a finite difference simulation, is shown for comparison. }
  \label{fig:1DTOV}
\end{figure*}

\begin{figure*}[t]
  \centering
  \includegraphics[width=\textwidth]{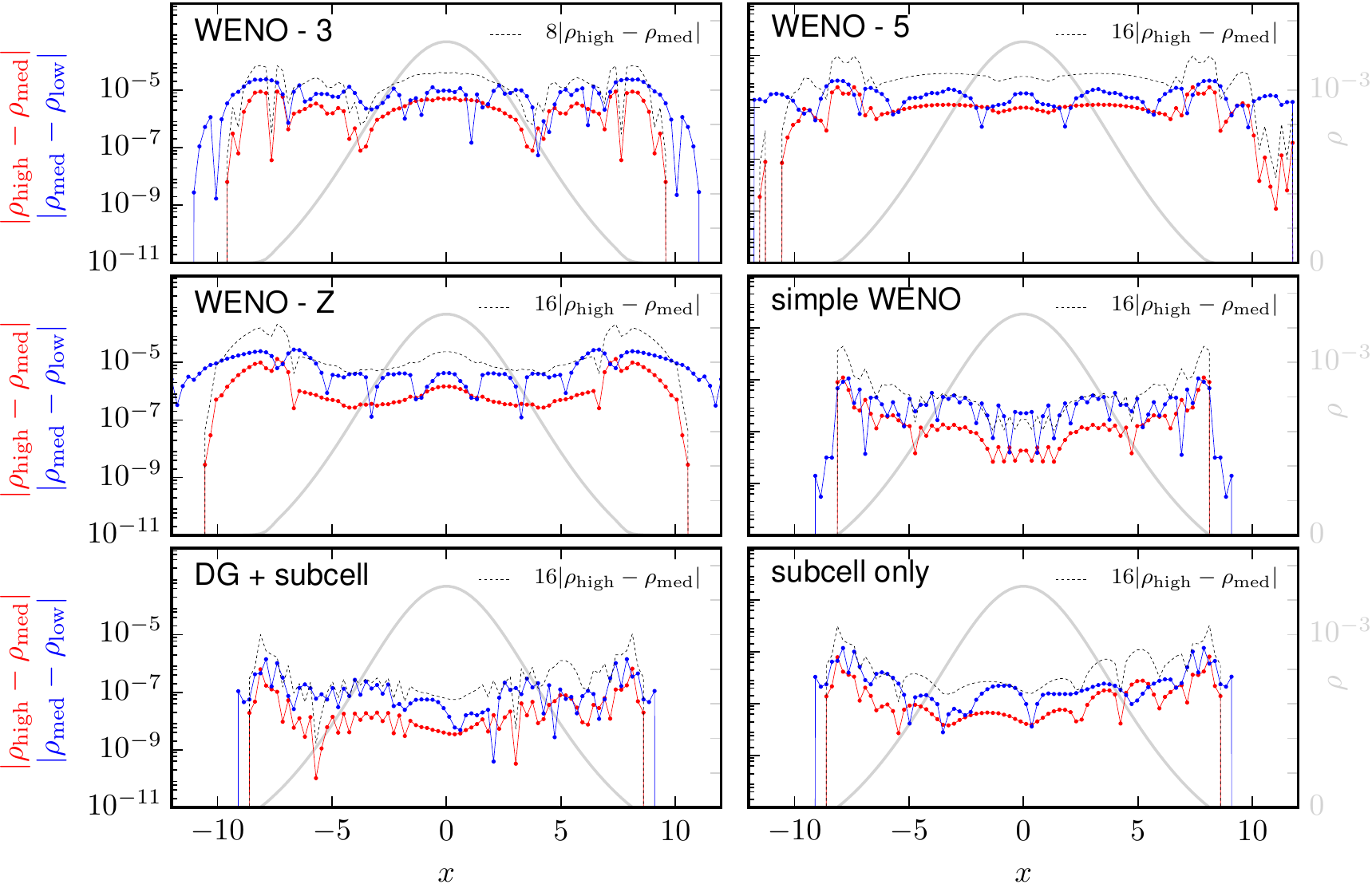}
  \caption{Convergence test for a 1D TOV star at $t=100$ for three resolutions $n_\text{high}=100,n_\text{mid}=50,n_\text{low}=25$,
  polynomials of order $N=3$, CFL $= 0.25$, $f_{atm} = 1$\e{-8} and $f_{thr} = 100$ evolved with DG and several shock resolution methods.}
  \label{fig:1DTOV_conv}
\end{figure*}

As the final test of our new implementation, we consider relativistic material in 
a curved spacetime background and present results for a TOV-star in Cowling-approximation 
in 1D, 2D, and 3D. Notice however, that the 1D and 2D description is not identical to the 3D star. 
Being more specific, surfaces of constant densities correspond for the 1D test to planes, 
for the 2D test to cylindrical shells, for the 3D test to spherical shells; 
cmp.~discussion below.

%%%%%%%%%%%%%%%%%%%%%%%%%%%%%%%%%%%%%%%%%%%%%%%%%%%%%%%%%%%%%%%%%%%%%%%%%%%%%%%%%%%%%%%%%%%%%%%%%%
\subsection{Initial configuration}
\label{sec:TOV_ID}
%%%%%%%%%%%%%%%%%%%%%%%%%%%%%%%%%%%%%%%%%%%%%%%%%%%%%%%%%%%%%%%%%%%%%%%%%%%%%%%%%%%%%%%%%%%%%%%%%%

Initial configurations for a single spherical symmetric neutron star 
are obtained by solving the TOV equation~\cite{Tol39,OppVol39a}.
The four-metric for a TOV star is given by
\begin{equation}
 {\rm d}s^2= - e^{2 \phi} {\rm d}t^2 + \left( 1- \frac{2 m}{R} \right)^{-1} {\rm d}R^2+R^2 {\rm d}\Omega^2.
\end{equation}
To obtain $m(R),\phi(R)$, and the pressure $p(R)$, the TOV equations 
\begin{eqnarray}
 \frac{{\rm d} \rho}{{\rm d} R}& = & \left(\rho (1+\epsilon)+p\right) \frac{m+4 \pi r^3 p}{R (R-2m)}\cdot \frac{1}{\frac{{\rm d}p} {{\rm d} \rho}},  \\
 \frac{{\rm d} m}{{\rm d} R}& = & 4 \pi R^2 \rho (1+\epsilon),  \\
 \frac{{\rm d} \phi}{{\rm d} R}& = & \frac{ m + 4 \pi R^3 p}{R(R-2m)}, 
\end{eqnarray}
are solved with an explicit fourth order Runge-Kutta algorithm. 
As starting values $\rho(R=0)=\rho_{\rm central}, m(R=0)=0,$ and  $\phi(R=0)=0$ are specified
and the system is closed by the polytropic EOS; Eq.~\eqref{eq:EOS_poly}. 
Afterwards a coordinate transformation is performed to obtain the metric in isotropic coordinates, 
which we use for the evolution, because $\alpha$ and $\psi^4$ can easily be obtained from
this form.
This solution describes the spacetime of a static, spherically symmetric star. 
However, due to the discontinuity at the stars surface and truncation errors, the evolution 
is non-trivial. 

%%%%%%%%%%%%%%%%%%%%%%%%%%%%%%%%%%%%%%%%%%%%%%%%%%%%%%%%%%%%%%%%%%%%%%%%%%%%%%%%%%%%%%%%%%%%%%%%%%
\subsection{1D-TOV tests}
%%%%%%%%%%%%%%%%%%%%%%%%%%%%%%%%%%%%%%%%%%%%%%%%%%%%%%%%%%%%%%%%%%%%%%%%%%%%%%%%%%%%%%%%%%%%%%%%%%
\label{sec:1D_TOV}

Before we are going to investigate the performance of our newly implemented algorithms in 
full 3D-simulations, we want to consider configurations similar to the TOV-star 
in just one dimension. We do not use spherical polar coordinates and stay in our Cartesian coordinate framework.
Thus, all derivatives along the $y$- and $z$-direction are set to zero to achieve translation symmetry, i.e.~$\partial_y \mathbf{f}^y$ and $\partial_z \mathbf{f}^z$ in
Eq.~\eqref{eq:con_law} are zero and also all first derivatives in $\mathbf{S}$~\footnote{
Notice that no second derivatives are present in Eq.~\eqref{eq:con_law} and that due to the restriction to 
Cowling approximation also first and second derivatives present in the metric field equations do not affect the simulation.}.
Therefore, the obtained spacetime is different to a spherical symmetric TOV star. 
Nevertheless, it is still a valid testbed for our numerical scheme and with the 
restriction to a fixed spacetime background, the initial condition are in hydrodynamical 
equilibrium. 

Because of the smaller computational costs, we will discuss in detail 1D-TOV results for all 
reconstruction algorithms, in particular we study WENO-3, WENO-5, WENO-Z, simple WENO reconstruction, 
as well as a DG+subcell and a pure subcell method for comparison. 
We have set in all our tests $f_{\rm atm}=10^{-8}$ and $f_{\rm thr} = 10^2$.
Figure~\ref{fig:1DTOV} shows the density $\rho$ (blue), the velocity $v^x$ (red), 
and the difference $|\rho-\rho_{\rm anal}|$ (green), where $\rho_{\rm anal}$ refers to initial condition 
constructed according to Sec.~\ref{sec:TOV_ID}. 
All reconstruction algorithms lead to stable evolutions. 
In general we observe 3 regions of troubled cells, 
the left star surface, the maximum of the density, and the right star surface. 
During the evolution some troubled cells are activated or deactivated, which explains 
why for WENO-Z reconstruction at the presented time $t=1000$
the surfaces are not marked as troubled. 

We observe that WENO-3, WENO-5, WENO-Z perform worst, i.e.~large velocities 
are present at the stars' surface and $|\rho-\rho_{\rm anal}|$ is larger as for the other reconstruction mechanisms 
(notice the different y-scales for $v^x$ and $|\rho-\rho_{\rm anal}|$). 
The best results are obtained with the simple WENO and DG+subcell methods.
The total $L_1$ errors of $\rho$ for the given setup are $6.0$\e{-7} for simple WENO and $4.6$\e{-7} for DG+subcell method.
The pure subcell evolution performs as good as the DG+subcell method. 

The advantage of the simple WENO and subcell methods can be understood by considering the effectively higher resolution compared to 
the other schemes. In the standard WENO case, only the cell averages are used for the componentwise reconstruction, therefore the effective resolution 
drops depending on the employed polynomial order. 
In contrast, the simple WENO approach uses the full information of the polynomial inside the cell and additionally uses only three cells for the reconstruction, 
thus no significant performance loss is obtained and the simple WENO reconstruction is a factor $1.57$ slower than the standard WENO-3 approach 
(a factor $1.40$ slower than the standard WENO-5 approach). 
Finally, in the DG + subcell method points are added in problematic regions. 
Because of the additional computational effort due to the projection between top- and subcells and 
the larger number of points in the troubled cells, the algorithm is a factor of $\sim 1.67 $ slower than the standard 
WENO method. 
Although not noticeable for 1D setups, we encounter for higher dimensional setups a significantly larger amount of memory, 
i.e. a $\sim 2.7$ times higher memory load for 2D runs 
($\sim 4.8$ times higher for 3D) when subcells are activated compared to standard WENO-3 simulations.
Nevertheless the DG+subcell approach seems to be a valid choice for further development, while it allows 
(i) to reuse well-tested FD schemes in troubled regions, 
(ii) give the most accurate results due to an effectively higher resolution in troubled regions,
(iii) allows a speed up compared to the usually employed FD codes, because of 
      a more effective DG method in large parts of the numerical domain. 

\begin{figure}[t]
  \centering
  \includegraphics[width=0.5\textwidth]{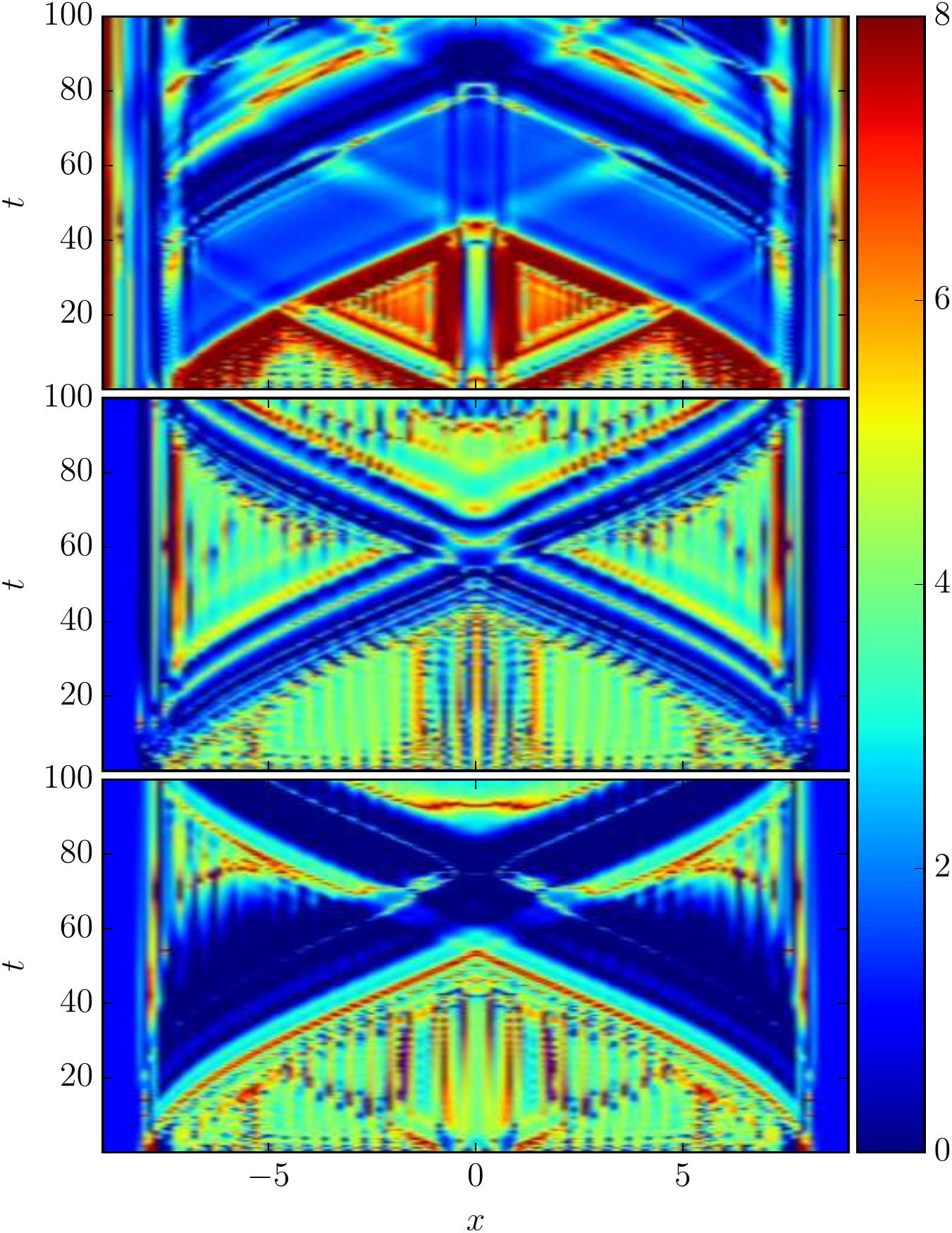}
  \caption{Convergence order for a 1D TOV star during evolution $t \in [0,100]$ for two resolutions $n_\text{high}=100,n_\text{low}=50$,
  polynomials of order $N=3$, CFL $= 0.25$, $f_{atm} = 1$\e{-8} and $f_{thr} = 100$ evolved with DG and three shock resolution methods: 
  standard WENO-3 (top), simple WENO (middle) and DG+subcell (bottom). 
  }
  \label{fig:1DTOV_conv_time}
\end{figure}

In Fig.~\ref{fig:1DTOV_conv} we present a pointwise convergence test for all methods. 
We compare evolutions with $25,50,100$ cells and use polynomials of order 3. 
The difference between the low and medium resolution is shown blue, while the difference between the medium and high resolution 
is shown red. We rescale the difference of the medium and high resolution according to the expected convergence order, 
i.e.~3rd order for WENO-3 and 4th order for the other schemes. 
We observe that in all cases we obtain roughly the expected convergence order. 
Furthermore in the logarithmic plots is clearly visible that for some setups the outer regions of the star is smeared out. 
In case of standard WENO algorithms larger stencils (WENO-5 and WENO-Z) lead to a numerical solution where the outer star layers 
are not fixed and no sharp surface is visible, this improves for the WENO-3 reconstruction. 
Contrary, the simple WENO and DG+subcell method keep the surface of the star fixed. 
In all runs higher resolution improves 
the results and less material is leaving the star. 

The simplicity of the 1D-TOV star allows us to consider setups with higher resolution 
than achievable in the corresponding 2D and 3D tests and
a more detailed analysis becomes possible. 
A recurring question is how regions with low order convergence
(because of low differentiability of the solution) affect regions
where the solution is smooth.  Specifically, do the regions of low
order remain localized, or if not, how quickly does the loss of
convergence spread through the entire domain? See for example \cite{KloWarHes11}, where
the wave equation with discontinuous initial data is studied, for which analytic results are
available in \cite{CocGuz08} 
predicting the growth of the non-convergent area with, e.g., the square-root of time,
$\sim\sqrt{t}$.

Figure~\ref{fig:1DTOV_conv_time} shows the convergence order during the first 
stages of the evolution for $50$ and $100$ cell setup. 
Presented are the WENO-3 (top panel), 
simple WENO (middle panel), and the subcell (bottom panel) 
evolution. For all panels, we observe that inside the star, 
where also cells are marked as troubled, the 
WENO-3 method shows $\sim$3rd order convergence and 
the simple WENO method a convergence order above 4. 
Furthermore, while for WENO-3 the error seems to corrupt the 
convergence in the entire star it seems to be localized for simple WENO 
for the entire simulation. 
For the subcell evolution we observe that the convergence order at the stars' center lies between second and third order, 
which is consistent with the employed flux methods implemented in the FD subcells~\cite{BerDie15}.
Artificially setting the center cells non-troubled cures this problem and leads locally to higher order convergence. However, it
has no influence on the global convergence order.
More problematic, a large error is traveling inwards from the outer surface for all simulations, which
leads to a lower convergence order for all setups. It is important to notice that this effect is 
not related to the movement of troubled cells. The region of troubled cells stays 
relatively fixed at the stars' surface. However, the flux across the cell surfaces seems to contain
lower order components. 
Regarding this fact, it is debatable whether one can obtain high order convergence in more general setups, 
e.g.~dynamical spacetimes and moving objects. 

%%%%%%%%%%%%%%%%%%%%%%%%%%%%%%%%%%%%%%%%%%%%%%%%%%%%%%%%%%%%%%%%%%%%%%%%%%%%%%%%%%%%%%%%%%%%%%%%%%
\subsection{2D TOV star}
%%%%%%%%%%%%%%%%%%%%%%%%%%%%%%%%%%%%%%%%%%%%%%%%%%%%%%%%%%%%%%%%%%%%%%%%%%%%%%%%%%%%%%%%%%%%%%%%%%

Considering the results of the previous section, the 
simple WENO and DG+subcell schemes seem to be preferable. 
However, we found, that the simple WENO method performs worse in higher dimensional problems as in the 1D case. 
Compared to the standard WENO reconstruction, where a smoother polynomial from several cell averages is constructed, 
the simple WENO methodology allows steeper gradients and has weaker smoothing influence. For runs of higher
dimensional problems we observe this smoothing to be crucial for the stability of the evolution. 
Furthermore, the simple WENO computation underlies a significant slowdown in $d>1$ Dimensions, because the
evaluation of the smoothness indicators is a quadratic form of all $(N+1)^d$ coefficients.
This is the reason why the standard WENO-3 scheme, which turns out to allow stable evolutions, is used instead.

\begin{figure}[t]
  \centering
  \includegraphics[width=0.49\textwidth]{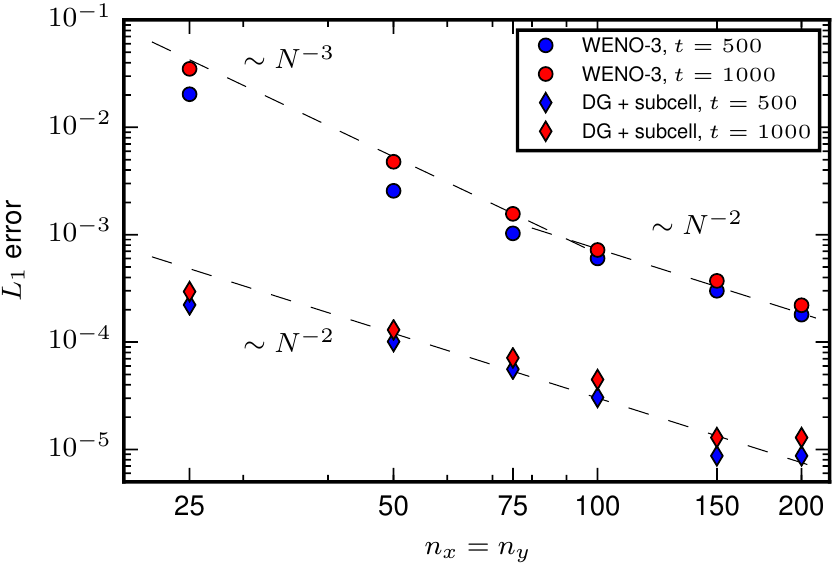}\\
  \caption{Density $L_1$ error for a 2D TOV star at $t=500$, $t=1000$ for six resolutions $n=25,50,75,100,150,200$,
  polynomials of order $N=3$, CFL $= 0.25$ and $f_{thr} = 100$ evolved with DG + standard WENO-3 and the DG + 
  subcell evolution method. For WENO-3 we set $f_{atm} = 1$\e{-8}, for the DG + subcell we set $f_{atm} = 1$\e{-9}. 
  The dashed black lines correspond to second / third order convergence.}
  \label{fig:TOV2D}
\end{figure}

We investigated the convergence of the two schemes, simple WENO and DG+subcell, for the 2D TOV star\footnote{
  As for the 1D test, we employ Cartesian coordinates and due to the restriction to a fixed spacetime background 
  also our 2D-TOV example is in hydrodynamical equilibrium.} 
regarding the density $L_1$ error. 
As shown in Fig.~\ref{fig:TOV2D}, we observe a convergence order of $\sim 2$ for the subcell scheme,
which indicates that the evolution error originating in the subcells spreads over the grid and leads to a
lower order of convergence. In comparison, the standard WENO-3 scheme converges in third order for coarse grids. 
The subcell method causes a much higher
memory load and longer calculation times (see Sec.~\ref{sec:1D_TOV}) because of the higher number of grid points in 
each direction. For this reason, we decided to use the standard WENO-3 method for the 3D simulation of a TOV star.

%%%%%%%%%%%%%%%%%%%%%%%%%%%%%%%%%%%%%%%%%%%%%%%%%%%%%%%%%%%%%%%%%%%%%%%%%%%%%%%%%%%%%%%%%%%%%%%%%%
\subsection{3D TOV star}
%%%%%%%%%%%%%%%%%%%%%%%%%%%%%%%%%%%%%%%%%%%%%%%%%%%%%%%%%%%%%%%%%%%%%%%%%%%%%%%%%%%%%%%%%%%%%%%%%%

\begin{figure}[t]
  \centering
  \includegraphics[width=0.49\textwidth]{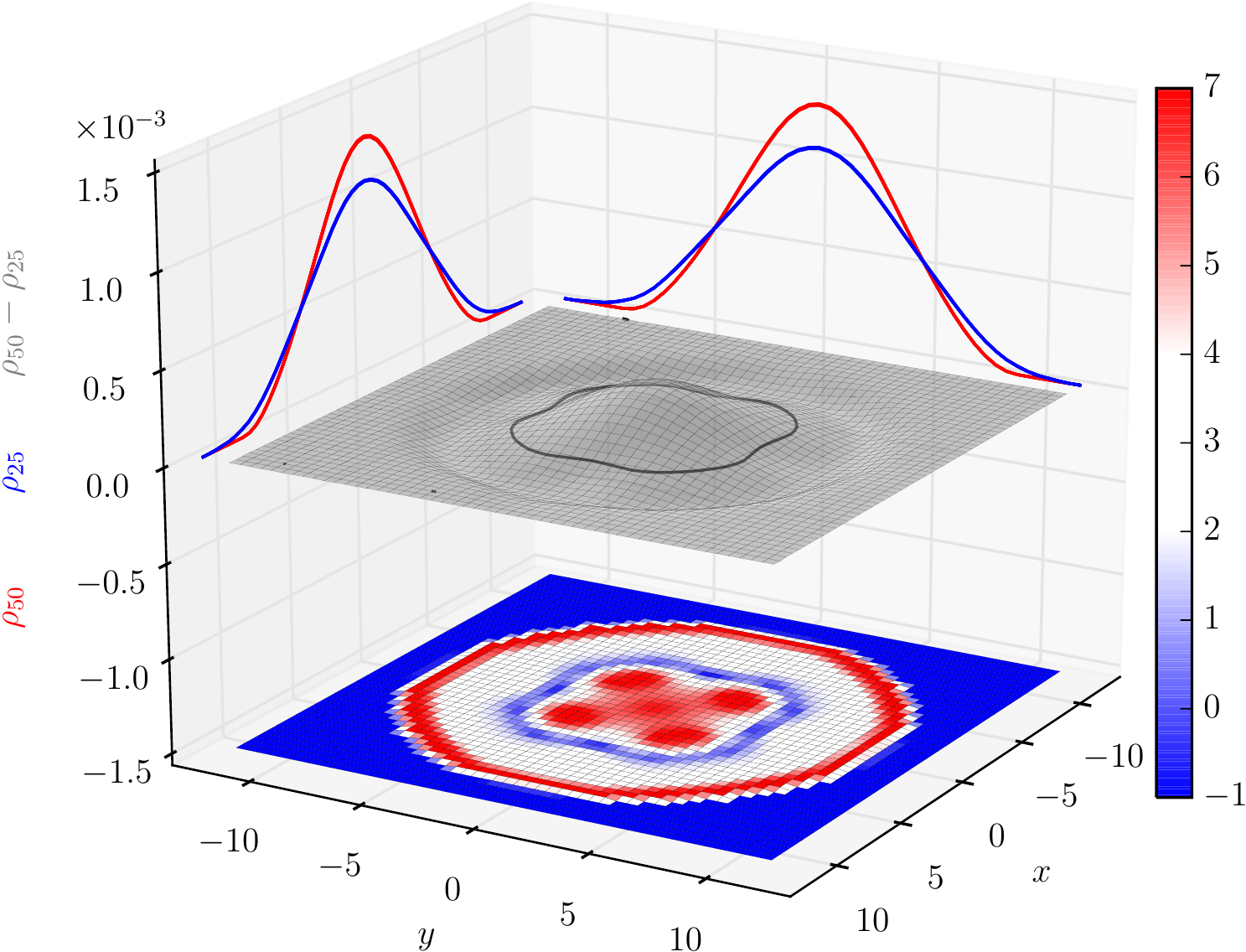}\\
  \caption{Pointwise convergence order in the $z=0$ plane for the density of a 3D TOV star at $t=500$ using 
  two resolutions $n_\text{high}=50$, $n_\text{low}=25$, polynomials of order $N=3$, CFL $= 0.25$, $f_{atm} = 1$\e{-8} 
  and $f_{thr} = 100$ evolved with DG and standard WENO-3. 
  The two density solutions for $n=50$ (red) and $n=25$ (blue) are shown on the axes $x=z=0$ and $y=z=0$. Their
  difference in the $z$-plane is shown in gray, the corresponding zero-crossing is indicated by the gray contour line.}
  \label{fig:TOV3D}
\end{figure}
\begin{figure}[t]
  \centering
  \includegraphics[width=0.49\textwidth]{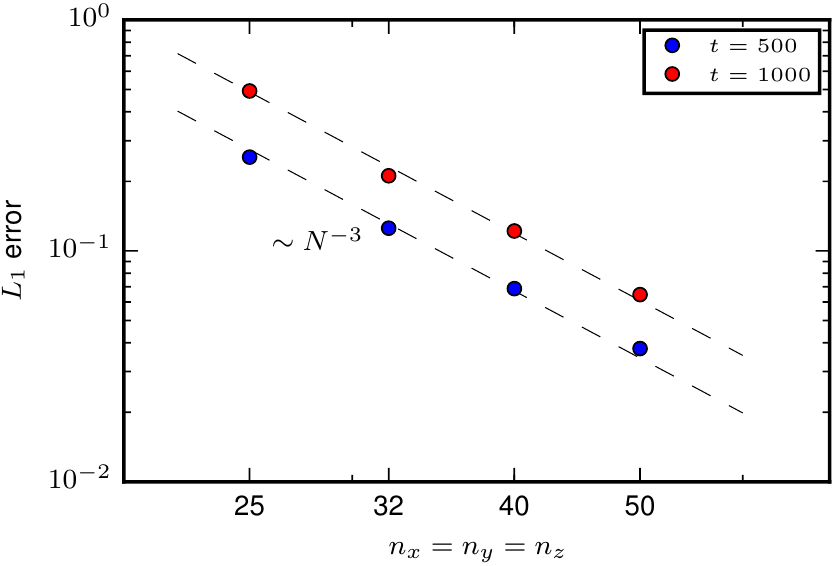}\\
  \caption{Density $L_1$ error for a 3D TOV star at $t=500$, $t=1000$ for four resolutions $n=20,32,40,50$,
  polynomials of order $N=3$, CFL $= 0.25$, $f_{atm} = 1$\e{-8} and $f_{thr} = 100$ evolved with DG and standard WENO-3. 
  The dashed black lines correspond to third order convergence.}
  \label{fig:TOV3D2}
\end{figure}

Considering a 3D TOV star, we are able to provide a stable simulation with a DG + standard WENO-3
method. Although we show our results up to $t=1000$, there is no evidence of any instabilities for
longer runs. 
We are considering two numerical setups at resolutions
$25 \times 25 \times 25$ and $50 \times 50 \times 50$ and the initial configuration as a reference solution. 
After a short transition, the numerical simulations reach an almost steady structure as shown in Fig.~\ref{fig:TOV3D}. 
The density profiles along the $x$- and $y$-axis are shown as red and blue lines. 
The difference between the densities for $z=0$ is presented as the gray shaded region. 
On the bottom panel, we present the computed convergence order. 
In large areas of the star second to fourth order convergence is present
and even higher convergence in its center and outside areas near the surface. The latter can be 
explained by the failure of the coarse grid setup to keep the density on atmosphere level outside
the star, whereas the fine grid setup does. 
The narrow band of low convergence (colored blue) inside the star 
can be explained as follows: The finer resolved solution stays closer
to the density maximum in the star center and zero at the star's surface, the opposite holds for the coarse resolution. 
Thus, the differences tend to zero, see solid black line, and the convergence drops locally.

As a global measurement of the convergence order, we present the $L_1$-norm for the 3D TOV star in Fig.~\ref{fig:TOV3D2}
for four different resolutions. Similar to the 2D test case, we observe an almost third order convergence (black dashed line) 
for the standard WENO-3 algorithm with third order polynomials. 
Using a fitting function of the form $A \cdot n_{x}^{-b}$ for the $L_1$-error, 
the obtained convergence order is $b=2.75$ for $t=500$ and $b=2.88$ for $t=1000$
and thus close to the theoretical expected value.

%%%%%%%%%%%%%%%%%%%%%%%%%%%%%%%%%%%%%%%%%%%%%%%%%%%%%%%%%%%%%%%%%%%%%%%%%%%%%%%%%%%%%%%%%%%%%%%%%%
%%%%%%%%%%%%%%%%%%%%%%%%%%%%%%%%%%%%%%%%%%%%%%%%%%%%%%%%%%%%%%%%%%%%%%%%%%%%%%%%%%%%%%%%%%%%%%%%%%
\section{Conclusion}
\label{sec:Conclusion}
%%%%%%%%%%%%%%%%%%%%%%%%%%%%%%%%%%%%%%%%%%%%%%%%%%%%%%%%%%%%%%%%%%%%%%%%%%%%%%%%%%%%%%%%%%%%%%%%%%
%%%%%%%%%%%%%%%%%%%%%%%%%%%%%%%%%%%%%%%%%%%%%%%%%%%%%%%%%%%%%%%%%%%%%%%%%%%%%%%%%%%%%%%%%%%%%%%%%%

In this work, we presented new algorithms implemented in the existing
\bamps code: a DG, a WENO-DG, and a mixed FD + DG-algorithm combined
with standard WENO~\cite{JiaShu96,QiuShu05} and a simple (compact)
WENO scheme~\cite{ZhoShu13}.  We tested all algorithms and
reconstruction methods with a number of tests starting with the
advection and Burgers equation, the main results being examples for
special and general relativistic hydrodynamics.  In almost all cases,
we were able to obtain the expected convergence order for smooth
solutions and also found a proper shock treatment in case of jumps and
discontinuities.

Our main result was the simulation of a single TOV-star, which we
modeled in the Cowling approximation, i.e.~for static geometric
variables.  In fact, while it has not been attempted yet to apply the
existing DG methods for the vacuum Einstein equations
\cite{Zum09,BroDieFie12} to 3D GRHD, we have demonstrated recently
that the pseudospectral multipatch method of \bamps \cite{HilWeyBru15}
works well for demanding 3D vacuum spacetimes (highly non-linear
gravitational waves that collapse to a black hole). The pseudospectral
penalty method developed in \cite{HilWeyBru15} can be viewed as a
special case of a DG method for the full Einstein equations in a
non-flux form. Furthermore, the present work on GRHD and the
wave-collapse simulations are compatible in the type of variables and
equations they use, and can run with the same spectral element grid
and polynomial basis functions (Chebyshev or Legendre Gauss-Lobatto
grids). Therefore, we do not expect any immediate obstacle to combine
the existing geometry code with the new GRHD methods.
 
One simplifying restriction in our implementation was the usage of a
simple troubled cell indicator.  We intend to study the influence of
different and more sophisticated troubled cell indicators. While our
simple setup allowed an easy implementation and stable evolutions, it
also marked maxima as troubled, which should be avoided in the future
application of the code.

Keeping the number of employed cells fixed, we found that for 1D
problems the subcell and simple WENO algorithms were the most accurate
ones. This can be easily understood, since the standard WENO method is
based only on cell averages for the reconstruction. In contrast, the
simple WENO method uses the knowledge of the entire polynomial, and
the subcell methods resolved troubled cells with effectively
$2N+1$-times more points.
In our examples, the simple WENO and subcell methods have some
drawback for higher dimensions.  Both methods come with a significant
overhead, and it is planned to investigate more efficient
implementations in the future. More importantly, the direct
application of the simple WENO reconstruction in 2D led to unexpected
instabilities, for example in the computation of the primitive from
the conservative variables, which one should be able to avoid. This
issue certainly deserves further study since the simple WENO method is
very promising based on the 1D results.

Due to the large computational cost of the subcell method, we only
employed the standard WENO reconstruction in 3D and investigated the
subcell method in 2D.  In our 2D examples, the subcell method turned
out to be (as expected) approximately second order.  For the standard
WENO-DG method we found 3rd order convergence for low and second order
convergence for high resolutions.  The observed third order
convergence is consistent with our results in the full 3D simulation.
However, for high number of cells we noticed that a higher than second
order convergence in the matter variables seems to be hard to obtain,
since (i) the computation of the $L_1$-norm of the error emphasizes
inaccurate, problematic regions, and (ii) errors propagate from the
surface of the star through the neutron star and ``corrupts'' the
order of convergence.  Although this fact can be seen as a setback, DG
methods allow a better parallelization and refinement strategy than
fixed FD codes and are one of the most promising methods to take into
account for future GRHD-code developments. 

Our work can be seen as a first step towards a complete 3D-DG
implementation for GRHD, since it employs DG-methods for GRHD problems
beyond the limitation of spherical symmetry as in previous work. It is
planned to further develop the numerical techniques by considering
adaptive mesh refinement, and to extend the physics to full general
relativity (beyond the Cowling approximation), which together will
allow numerical simulations of astrophysical systems consisting of
single and binary neutron stars.

\begin{acknowledgments}
  It is a pleasure to thank E~Harms, D.~Hilditch, N.~Moldenhauer, 
  M. Pilz, and H.~R\"uter for helpful discussions.
  This work was supported in part by DFG grant
  SFB/Transregio~7 ``Gravitational Wave Astronomy'' and the
  Graduierten-Akademie Jena.
  The authors acknowledge the usage of computer resources at 
  the GCS Supercomputer SuperMUC, 
  JUROPA at J\"ulich Supercomputing Centre, and 
  at the Institute of Theoretical
  Physics of the University of Jena.
  
\end{acknowledgments}

\appendix

\section{DG method for scalar conservation laws in 1D}
\label{sec:appendixDG}
\label{app:once_twice}

Following \cite{Kop09,HesWar08}, we summarize some aspects of the DG method
that already arise in the non-linear, scalar, one-dimensional case. We add
some details relevant to the present work concerning implementation issues
and the equivalence of the once and twice integrated form of the equations.
One of our goals is the combination of the DG method for relativistic matter
with the pseudospectral penalty method of \cite{HilWeyBru15} for the geometry,
which is not using a flux conservative form, but is close to the strong formulation
of the DG method given below in (\ref{Mdeltu2}). Therefore we examine the question
how the discretized equations for the weak and strong form are related.

%%%%%%%%%%%%%%%%%%%%%%%%%%%%%%%%%%%%%%%%%%%%%%%%%%%
\subsection{Derivation of the discretized equations}

Consider
\beq
\partial_t u + \partial_x f(u) = 0
\label{vdelu0}
\eeq
for a function $u(t,x)$ and a flux function $f(u(t,x))$ on
the interval $I=[-1,1]$. Given a space of test functions on $I$, we obtain the weak
form of the conservation law by integration. For a test function $v(x)$,
\beq
(v,\partial_t u) + (v, \partial_x f) = 0.
\label{vdelu}
\eeq
Later we assume that the scalar product of two functions is
$(f_1,f_2)=\int_{-1}^1 f_1(x)f_2(x)dx$, i.e.\ we assume the trivial
measure which is the natural weight for the polynomial basis of
Legendre polynomials.

Part of the DG method is a special treatment of the flux at the boundaries of $I$,
where we replace the flux $f$ by a non-unique choice of a numerical flux $f^*$.
Integrating (\ref{vdelu}) by parts in space we arrive at two versions
of the conservation law,
\bea
  (v,\partial_t u) - (f, \partial_x v) &=& - [v f^*],
\label{vdelu1}
\\
  (v,\partial_t u) + (v, \partial_x f) &=& [v (f-f^*)],
\label{vdelu2}
\eea
where $[g]= g(1)-g(-1)$ for any function $g(x)$.
Eqn.\ (\ref{vdelu1}) is obtained by integrating by parts and replacing
$f$ by $f^*$ at the boundary. Eqn.\ (\ref{vdelu2}) is obtained from
(\ref{vdelu1}) by integrating by parts once more but leaving the
resulting boundary term unchanged.  We refer to (\ref{vdelu1}) and
(\ref{vdelu2}) as the weak and strong form, respectively, or as the
once and twice integrated flux equation to avoid confusion with the
original ``strong'' form, (\ref{vdelu0}).

The nodal DG spectral element method is based on a choice of $N+1$
distinct nodes $x_i\in I$. Such nodes define the unique $N$-th-order
Lagrange polynomials $\ell_i(x)$, for which $\ell_i(x_j)=\delta_{ij}$.
We choose the $x_i$ to be the collocation points for Legendre-Gauss (LG) or 
Legendre-Gauss-Lobatto (LGL) integration. This allows the approximation
of $u(x)$ by a nodal expansion that has the interpolation property
$u(x_i)=u_i$, $I_mu(x) = \sum_{i=0}^N u_i \ell_i(x)$.

When the meaning is clear from context, we write $u$ instead of $I_mu$. 
A key feature of the nodal expansion is that it works equally well for linear and
non-linear functions, in particular
\beq
  u(x) = \sum_{i=0}^N u_i \ell_i(x), 
\quad
  f(u(x)) = \sum_{i=0}^N f_i \ell_i(x),
\label{nodaluf}
\eeq
where $f_i=f(u_i)=f(u(x_i))$.

The nodal approximation with $N$-th-order polynomials leads to 
discretized versions
of the conservation laws (\ref{vdelu1}) and
(\ref{vdelu2}). Choose test functions $v(x)=\ell_i(x)$, and insert (\ref{nodaluf})
to obtain
\bea
  M \partial_t u - S^T f &=& -[\ell f^*],
\label{Mdeltu1}
\\
  M \partial_t u + S f   &=& [\ell (f-f^*)],
\label{Mdeltu2}
\eea
where we have introduced the mass matrix $M$ and stiffness matrix $S$,
\beq
   M_{ij} = (\ell_i,\ell_j),
\qquad
   S_{ij} = (\ell_i,\partial_x\ell_j).
\eeq
We use matrix notation and a summation convention, 
e.g.\ $S f\equiv S_{ij}f_j\equiv \sum_{j=0}^N S_{ij}f_j$.

An important point is that in general the mass matrix is not diagonal,
that is, the characteristic Lagrange polynomials are not necessarily orthogonal.
Specifically, for LGL the mass matrix is not diagonal, while for LG 
it is diagonal. However, for both LGL and LG the matrix is symmetric and
invertible.
The stiffness matrix is directly related to the derivative matrix, 
\beq
   D_{ij} = \partial_x\ell_j(x_i),
\eeq
which approximates the pseudospectral derivative at the nodes
by $(\partial_x u)(x_i)=D_{ij} u_j$. We have
\cite{HesWar08}
\beq
   S = M D,
\quad
   M^{-1}S = D, 
\quad 
   M^{-1}S^T = M^{-1}D^TM.
\eeq
%where we used $M=M^T$.
%
Given $M$, $D$, and a prescription for $f^*$, we solve 
the explicit time-integration problem based on (\ref{Mdeltu1}) or
(\ref{Mdeltu2}),
\bea
\partial_t u - (M^{-1}D^TM) f &=& - M^{-1}[\ell f^*],
\label{deltu1}
\\
\partial_t u + D f &=& M^{-1}[\ell (f-f^*)],
\label{deltu2}
\eea
for the descretized, time-dependent function values $u_i(t)$.  

The method generalizes immediately to a partition of any interval
$[a,b]\in\mathbb{R}$ into several elements $I_j$ with an appropriate
mapping of the coordinates and with a coupling of neighboring elements
through $f^*$.

%%%%%%%%%%%%%%%%%%%%%%%%%%%%%%%%%%%%%%%%%%%%%%%%%%%
\subsection{Implementation issues}

Let us comment on some implementation issues, specifically for the LGL method.
The nodes $x_i$ and the LGL integration weights $w_i$ are obtained from
the Legendre polynomials, for which simple but stable and accurate algorithms are
available, e.g.\ \cite{Kop09}. The nodes $x_i$ are the $N-1$ roots of 
$\partial_xP_{N}(x)$ combined with the endpoints of the interval, $-1$ and $1$,
for a total of $N+1$ nodes. The integration weights are
$w_i = 2 / (N (N+1) P_N(x_i)^2)$.
Various other quantities are determined without further reference to
the Legendre polynomials by general formulas for Lagrange
interpolation. The weights for barycentric interpolation are
$c_i= \prod_{j=0,j\neq i}^N 1/(x_i-x_j)$. The derivative matrix is
\bea
   D_{ij} &=& \partial_x\ell_j(x_i) = \frac{c_j}{c_i}\frac{1}{x_i-x_j},
\quad i\ne j,
\label{Dij}
\\
   D_{ii} &=& - \sum_{j=0}^N D_{ij}.
\eea
The equation for the diagonal term ensures that the numerical derivative of
a constant like $u_i=1$ is zero \cite{BalTru03}.
Since the endpoints are included among the nodes, $x_0=-1$ and $x_N=1$,
\beq
[\ell_i g]_{-1}^1 = \ell_i(1) g(1) - \ell_i(-1) g(-1)
= \delta_{iN} g_N - \delta_{i0} g_0.
\eeq

There are several ways to compute the mass matrix
$M_{ij}=(\ell_i,\ell_j)$.  One option is to perform the integration
numerically according to the Gauss formula associated with the nodes,
which approximates the integral of a function
$g(x)$ using the integration weights $w_i$,
\beq
  \int_{-1}^1 g(x) dx \simeq \sum_{i=0}^N w_i g(x_i).
\label{Gaussint}
\eeq
This integration is exact if $g(x)$ is a polynomial of degree up to $2N+1$ for LG
and up to $2N-1$ for LGL. Since the integrand $\ell_i\ell_j$ for the mass matrix 
is of degree $2N$, for LG the numerical integral is exact,
\beq
  M_{ij}=(\ell_i,\ell_j) = (\ell_i,\ell_j)_N = w_i \delta_{ij},
\eeq
where $(f,g)_N = \sum_i w_i f_i g_i$ denotes the numerical scalar product.
However, for LGL we only obtain the approximation
\beq
  M_{ij}=(\ell_i,\ell_j) \simeq (\ell_i,\ell_j)_N = w_i \delta_{ij}.
\label{Mapprox}
\eeq
It turns out that this approximation, also called mass lumping,
is equivalent to a certain filter
that strongly affects the highest mode in the Legendre basis and that can
reduce the effective order of the approximation \cite{GasKop11}. 
In the context of spectral element methods of comparatively high order, say 
$N=10$,
approximating $M$ for LGL by the diagonal matrix as in (\ref{Mapprox}) 
is considered standard in \cite{Kop09}. However, for orders around 
$N=2,3,4$, it is often preferable to evaluate $M_{ij}=(\ell_i,\ell_j)$ for
LGL without approximation \cite{HesWar08}. For example \cite{HesWar08,GasKop11},
$M^{-1} = V V^T$,

where $V$ is the generalized Vandermonde matrix for the normalized Legendre
polynomials. This relation follows from the expansion of the Legendre polynomials
in the Lagrange basis. Computing the difference to the diagonal approximation 
we find for LGL
\beq
  M^{-1}_{ij} = \frac{1}{w_i}\delta_{ij} + \frac{N+1}{2} P_N(x_i) P_N(x_j).
\label{myMinv}
\eeq
Alternatively, note that $M$ can also be computed directly as the
analytic integral $(\ell_i,\ell_j)$, either by term by term
integration after expanding the product $\ell_i(x)\ell_j(x)$, or by
exact Gauss integration on a secondary grid with $N+2$ points. (In
experiments, $N+3$ gives somewhat more accurate results.)  However, we
still have to find the inverse of $M$ numerically. For large $N$,
(\ref{myMinv}) may be preferred.

%%%%%%%%%%%%%%%%%%%%%%%%%%%%%%%%%%%%%%%%%%%%%%%%%%%
\subsection{Equivalence of once and twice integrated forms}

For the continuum problem, we perform the integration by parts
\beq
   (v,\partial_xf) = [v f] - (f,\partial_xv).
\eeq 
Under specific but quite general conditions the discretized
equations satisfy the corresponding summation by parts property exactly.
In this case the once and
twice integrated DG methods are numerically identical. There may be
round-off errors, but there are no systematic errors that only
converge away with increasing $N$. This is fully explained in 
\cite{CarGot96,GasKop11,Gas13}. 

Given the present setup, it is straightforward to show algebraic equivalence of 
(\ref{Mdeltu1}) and (\ref{Mdeltu2}).
The difference between those two equations is
\bea
   S f &=& [\ell f] - S^T f,
\\
   S &=& [\ell \ell] - S^T,
\label{Sfsum}
\eea
for all $f_i$, and independently of the choice of $f^*$ or the computation of $M$. 
In the transition to (\ref{Sfsum}) we use that $f$ is approximated
by an $N$-th-order polynomial, (\ref{nodaluf}). By definition of $S_{ij}$,
\bea
 S_{ij} 
 &=& (\ell_i,\partial_x\ell_j)
\nonumber
\\
 &=& [\ell_i \ell_j] - (\ell_j, \partial_x\ell_i)
\nonumber
\\
 &=& [\ell_i \ell_j] - S_{ji}, 
\eea
so the summation by parts property (\ref{Sfsum}) does indeed hold.
Summation by parts is exact for LG and LGL even if $S_{ij}$ is
defined by numerical integration because
\beq
S_{ij} = (\ell_i,\partial_x\ell_j) = (\ell_i,\partial_x\ell_j)_N
\label{Sellell}
\eeq
since $\ell_i\partial_x\ell_j$ is a polynomial of degree $2N-1$.

It is instructive to make the summation by parts formula for the LGL method 
more explicit. From (\ref{Sellell}) and (\ref{Gaussint}),
\bea
  S_{ij} &=& \sum_k w_k \ell_i(x_k) \partial_x\ell_j(x_k)
 = w_i D_{ij}.
\eea
(Incidentally, this means that $S_{ij}=M_{ik}D_{kj}= w_i D_{ij}$ for both LG and LGL).
Hence (\ref{Sfsum}) becomes
\beq
  w_i D_{ij} = [\ell_i\ell_j] - w_j D_{ji}. 
\label{wDllwD}
\eeq
We now restrict ourselves to the LGL case.
A priori it is not clear how a simple rescaling and a transpose of the 
derivative matrix leads to the term 
$[\ell_i\ell_j] = (\delta_{iN}-\delta_{i0})\delta_{ij}$,
which is a diagonal matrix with
non-vanishing entries only in two of the corners.
For $i=j$,
\bea
  D_{ii} &=& \partial_x l_i(x_i) = \frac{1}{2w_i}(\delta_{iN} -
  \delta_{i0}),
\\
  2 w_i D_{ii} &=& \delta_{iN} - \delta_{i0} = [\ell_i\ell_i],
\eea
so (\ref{wDllwD}) 
is satisfied on the diagonal. 
In particular, we see how the boundary terms come about.
For $i\neq j$, (\ref{wDllwD}) becomes
\beq
   D_{ij} = - \frac{w_j}{w_i}(D^T)_{ij},
\eeq
from which we obtain with (\ref{Dij}) that
\beq
   \frac{c_j}{c_i}=\frac{w_j}{w_i}\frac{c_i}{c_j}.
\eeq
In other words, the summation by parts rule implies for LGL points
a relation between the integration weights $w_i$ and the barycentric
interpolation weights $c_i$,
\beq
   (c_i^{LGL})^2 = C w^{LGL}_i ,
\label{mycofw}
\eeq
for some constant $C$. 
Surprisingly, the explicit relation between $w_i^{LGL}$ and $c_i^{LGL}$ was only
found recently, see \cite{WanHuyVan12} 
%[\texttt{http://arxiv.org/abs/1202.0154}] 
on such relations for Jacobi polynomials. For our case,
\beq
   c_i^{LGL} = C_N (-1)^i \sqrt{w_i^{LGL}},
\label{cofw}
\eeq
where $C_N$ is an explicitly known constant that depends on the
number of points. In summary, for LGL (or
analogously for LG), we can start from the general result on summation
by parts and arrive at a partial proof of (\ref{cofw}), or we can
start from relations like (\ref{cofw}) and prove summation by parts
without directly using partial integration in the continuum.
    
\bibliography{paper20150828.bbl}

\end{document}